\documentclass[a4paper,twocolumn,11pt,unpublished]{quantumarticle}
\pdfoutput=1
\usepackage[utf8]{inputenc}
\usepackage[english]{babel}
\usepackage[T1]{fontenc}
\usepackage{amsmath}
\usepackage{graphicx}
\usepackage{tikz}
\usepackage{subcaption}
\usepackage{braket}
\usepackage{lipsum}
\usepackage[numbers]{natbib}
\usepackage[hyphens]{url}
\usepackage{hyperref}

\makeatletter
\providecommand{\@afterenddocumenthook}{}
\makeatother

\begin{document}

\title{Sensitivity Evaluation of SU(1,1) Interferometers with Arbitrary Input Probe State and Homodyne Detections}

\author{Sonu Jana}
\email{sonujana@kgpian.iitkgp.ac.in}
\thanks{}
\affiliation{Centre for Interdisciplinary and Convergent Technologies, Indian Institute of Technology Kharagpur, Kharagpur 721302, West Bengal, India}

\author{Dhruv Baheti}
\affiliation{Department of Physics, Indian Institute of Technology Kharagpur, Kharagpur 721302, West Bengal, India}

\author{Paul Grossiord}
\affiliation{Universit\'e Paris-Saclay, CNRS, CentraleSup\'elec, LuMIn, Orsay, France}

\author{Fabien Bretenaker}
\affiliation{Universit\'e Paris-Saclay, CNRS, CentraleSup\'elec, LuMIn, Orsay, France}
\affiliation{Centre for Interdisciplinary and Convergent Technologies, Indian Institute of Technology Kharagpur, Kharagpur 721302, West Bengal, India}

\author{Nadia Belabas*}
\affiliation{Centre for Nanosciences and Nanotechnologies, CNRS, Universit\'e Paris-Saclay, UMR 9001, 10 Boulevard Thomas Gobert, 91120 Palaiseau, France}

\author{Syamsundar De*}
\email{syamsundarde@atdc.iitkgp.ac.in}
\affiliation{Centre for Interdisciplinary and Convergent Technologies, Indian Institute of Technology Kharagpur, Kharagpur 721302, West Bengal, India}
\maketitle
 
\begin{abstract}
We provide a general theoretical derivation of the phase sensitivity achieved by SU(1,1) interferometers under homodyne detection. The general expressions obtained accommodate arbitrary input states and include internal and external losses. In this systematic review, both full SU(1,1) interferometers with two parametric amplifiers and the truncated interferometers with only one parametric amplifier are examined. We investigate scenarios involving both single-output ports and joint homodyne detection, and consider parametric amplifiers with equal gains or with a boosted gain second amplifier. Our analytical formulation provides physical insight and understanding of the improvements in the sensitivity, which are shown to originate from noise reduction and/or signal amplification, depending on the configurations and practical implementations. Surprisingly, the configuration with single-output mode detection and parametric amplifiers with equal gains exhibits the highest robustness to very high internal losses. We finally apply this framework to a ubiquitous $\ket{\alpha} \otimes \ket{0}$ input two-mode coherent probe state. This approach permits the comparison of different strategies and the optimization of the interferometer performance in the presence of losses. In particular, we determine which amplification and detection configurations provide the best performance, depending on the level of losses. This exemplifies how this general analytical approach provides a powerful tool to design quantum-enhanced interferometers and achieve optimal sensitivity with selected probe states and homodyne detection. 
\end{abstract}

\section{Introduction}

    Interferometry is a foundational tool in precision metrology, enabling high-sensitivity measurement of physical quantities that induce phase shifts between optical paths \cite{1_michelson1887relative,2_de2019review,3_yang2018review,4_jaros2022fiber}. Notable examples include gravitational wave detection using laser interferometers \cite{5_abbott2016observation}. However, the phase sensitivity of standard interferometers is fundamentally constrained by vacuum fluctuations entering the unused port, leading to the shot-noise-limited scaling of $\frac{1}{\sqrt{N}}$, where $N$ is the average number of phase-sensing photons  \cite{6_takeoka2017fundamental,7_caves1981quantum}. This Standard Quantum Limit (SQL) can be surpassed using non-classical states of light such as squeezed vacuum or NOON states, enabling scaling improvements up to the Heisenberg limit \cite{8_tan2014enhanced,9_dowling2008quantum,10_luca2008mach}.
    
    A complementary strategy that does not require the input of exotic quantum states, which are challenging to prepare, involves modifying the interferometer architecture itself. In the SU(1,1) interferometer, originally proposed by Yurke et al. \cite{11_yurke19862}, beam splitters are replaced with optical parametric amplifiers (PAs), which generate non-classical light via nonlinear wave mixing. This design enables phase-sensitive amplification and can surpass the SQL without requiring externally prepared quantum probe states. A practical variant was introduced by Plick et al. \cite{12_plick2010coherent}, using coherent and vacuum input states to produce stronger output signals and stimulate renewed interest in nonlinear interferometry \cite{13_lukens2018broadband,14_liu2018loss,15_liu2019optimum,16_manceau2017detection,Seyfarth2020wignerfunction}. Since then, a variety of SU(1,1) configurations (truncated SU(1,1), nested SU(2) inside SU(1,1), displacement-assisted SU(1,1), etc.), input states (vacuum-vacuum, coherent-vacuum, coherent-coherent, coherent-squeezed vacuum, etc.), and detection strategies have been explored, including photon counting, parity measurement, and homodyne detection \cite{17_szigeti2017pumped,18_du20202,19_kang2024phase,20_Kumar2025,21_ou2020quantum,22_9571144,23_li2014phase,24_wang20211,Ferreri2021spectrallymultimode}.
    
    Optimising phase sensitivity in such systems involves selecting appropriate input state preparation and measurement strategy. While the quantum Cramer-Rao bound (QCRB) is claimed to provide the ultimate sensitivity limit for a given probe state \cite{25_you2019conclusive,26_jarzyna2012quantum,27_anderson2017optimal,Kranias2025metrological}, it does not prescribe the specific measurement to perform to reach it. In this work, we adopt a complementary approach: for a fixed measurement scheme, namely homodyne detection, we develop a generalised theoretical framework to compute the phase sensitivity of SU(1,1) interferometers for \textit{arbitrary two-mode input states}. We focus here on homodyne detections; an analogous formulation for photon-number measurement will be presented in a forthcoming work. Our method links the average values and variances of the detected quadratures to the input ones, independently of the input field states. Unlike previous analyses limited to specific state choices \cite{12_plick2010coherent,21_ou2020quantum,22_9571144,23_li2014phase}, our formalism provides a complete analytical expression for phase sensitivity that remains valid even in the presence of both internal (i.e., inside the interferometer) and external (i.e., before the detection) losses. This analytical generality allows a unified treatment of diverse input configurations within a single formalism, thereby connecting and extending previously state-specific analyses. We thus provide a deeper physical insight into the sensitivity enhancement mechanisms that was not accessible in earlier state-dependent approaches.
    
\begin{figure}[htp]
    \centering
    \includegraphics[width=1\linewidth]{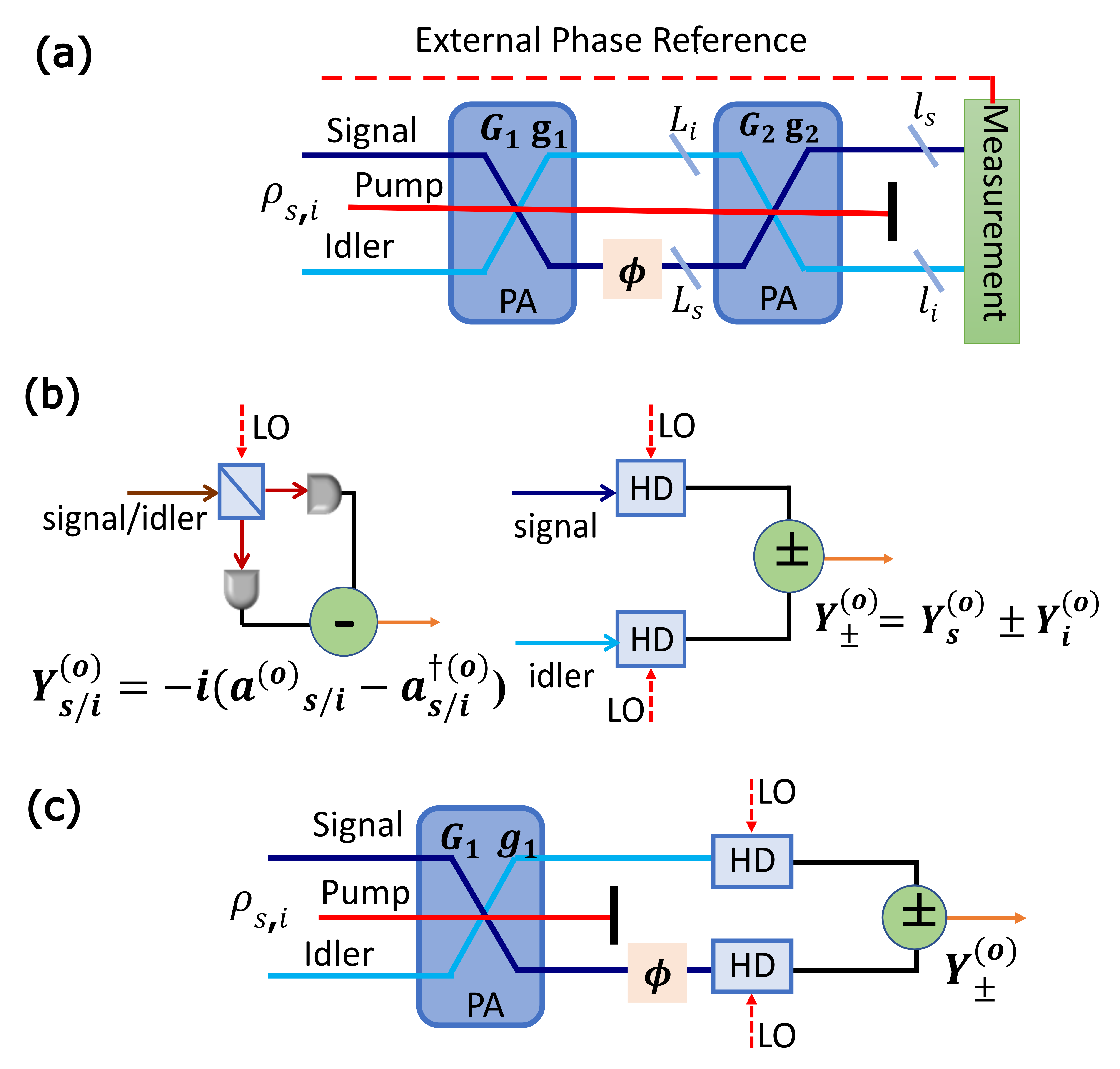} 
    \label{fig:1}
    
    \caption{\label{Fig01} \textbf{(a)} Schematic of a full SU(1,1) interferometer. The beam splitters of a conventional Mach-Zehnder interferometer are replaced by parametric amplifiers (PAs). A strong pump drives the first PA, which mixes the two modes (\textcolor{blue}{signal (s)} and \textcolor{cyan}{idler (i)}) of an arbitrary two-mode input state $\rho_{s,i}$. After the first PA (characterized by $g_1$ and $G_1$ see main text), a phase shift $\phi=\phi_0+\delta\phi$ is applied to one mode (\textcolor{blue}{signal}) relative to the pump and the other mode (\textcolor{cyan}{idler}). Both modes, together with the pump, then act as inputs to the second PA (characterized by $g_2$ and $G_2$). Internal and external losses are modeled by fictitious beam splitters placed inside and outside the interferometer (losses $L_s$, $L_i$, $l_s$, and $l_i$). Measurements are performed at the outputs of the second PA. \textbf{(b)} Different homodyne measurement schemes. Left: Single-port homodyne detection (HD): homodyne measurement ($Y^{(o)}_{s/i}$) is performed on one output port (\textcolor{blue}{signal (s)} or \textcolor{cyan}{idler (i)}) while the other port (\textcolor{cyan}{idler} or \textcolor{blue}{signal}) remains unused. Right: Joint homodyne detection: homodyne measurements are performed on both output ports, and the resulting photocurrents are electronically combined (added or subtracted ($\pm$)) to measure $Y^{(o)}_{\pm}$. LO: Local oscillator \textbf{(c)} Schematic of the truncated SU(1,1) interferometer. The setup has only one parametric amplifier. A joint homodyne detection ($Y^{(o)}_{\pm}$) is performed directly on the two output modes of the PA. }
    \end{figure}

    To make the present analysis as general as possible, we consider both single-mode homodyne detection and combined homodyne detection of the idler and signal modes at the interferometer output. Moreover, we consider three different SU(1,1) interferometer configurations: the so-called \textit{balanced} one in which the gains of the two parametric amplifiers are equal, the \textit{unbalanced configuration} in which the gain of the second parametric amplifier is much larger than the gain of the first one, and finally the so-called \textit{truncated} SU(1,1) interferometer \cite{28_anderson2017phase,29_gupta2018optimized} that consists in a single parametric amplifier and a phase detection via a joint homodyne detection (Fig.\,\ref{Fig01}(b)). The influence of the losses is considered in all three balanced, unbalanced, and truncated cases.
    
    Previous studies have shown that, for a $\ket{\alpha} \otimes \ket{0}$ input, (here $|\alpha|^2$ is the average photon number of the coherent sate injected into one arm of the interferometer, while the other arm receives a vacuum state) the full SU(1,1) interferometer is equivalent to its truncated version under joint homodyne detection \cite{28_anderson2017phase,29_gupta2018optimized}. We generalize this result and show that the equivalence holds for any arbitrary input state. While both full and truncated interferometer configurations yield identical phase sensitivity, we show that the underlying mechanisms for sensitivity enhancement differ. We clarify that in the full setup, sensitivity is enhanced either through signal amplification or noise suppression, depending on the operating point of the interferometer (respectively at the dark fringe and at the bright fringe, regardless of the input state). In contrast, the truncated configuration benefits from simultaneous signal enhancement and noise reduction at both the dark and bright fringes.
    
    Finally, the predictive power of our framework is illustrated in Section \ref{Application}. We analyse the phase sensitivity achieved by harnessing a $\ket{\alpha}\otimes\ket{0}$ input state in the presence of internal and external losses. Although this configuration has been extensively studied in both lossless and lossy regimes \cite{13_lukens2018broadband,15_liu2019optimum,16_manceau2017detection,21_ou2020quantum,30_hudelist2014quantum,31_marino2012effect,32_ou2012enhancement}, a complete analytical treatment covering all forms of loss and various homodyne detection schemes was still missing. Using a fully analytical formalism, we not only reproduce the known results but also obtain new physical insights into the behavior of the interferometer under realistic conditions. The readers interested in the practical optimization of their interferometers probed by coherent states in the presence of losses can thus directly refer to Section \ref{Application}. Notably, Fig.\,\ref{Fig05} provides a strategy to determine which gain and measurement configurations optimize the sensitivity for a given amount of losses. In particular, we show that under relatively high internal losses,  single-port homodyne detection with balanced gains for the two amplifiers outperforms the unbalanced gain configuration with single-port homodyne or joint (dual-port) homodyne detections. We further demonstrate that single-port homodyne detection exhibits a robustness to a wide range of internal losses and maintains a phase-shift estimation sensitivity beyond that of a classical, lossless Mach–Zehnder interferometer. These results highlight the practical robustness of our approach and its relevance for real-world quantum sensing applications.

\section{Generalized probe state formalism}
\label{probe formalism}

    The standard protocol for interferometric phase-shift estimation involves the unitary evolution of an input quantum probe state under a Hamiltonian that depends on the phase shift, followed by a measurement of the state that underwent the interferometer transformation using a suitable Hermitian operator.
    In our analysis, we do not restrict ourselves to any specific input probe state. We consider a general two-mode optical input state $\rho_{s,i}$, where $s$ and $i$ denote the input signal and idler modes. Sensitivity expressions for the parameter of interest, the interferometric phase shift, are derived for both single and joint homodyne (field quadrature) detection at the output. 
    We adopt the Heisenberg representation to propagate the field operators through the interferometer. This approach isolates the state-dependent contributions to the sensitivity, allowing one to select any input statistics and feed them into a general formula. The evolution of observables depends solely on the interferometer configuration and remains independent of the specific input state.

\subsection{Interferometer transformation}
    Consider the interferometer shown in Fig.\,\ref{Fig01}(a). The interferometer transformation is obtained by applying the individual transformations of each element in the right order. The PA transformation is given as an SU(1,1) transformation coupling the signal and idler input and output operators according to:
    \begin{equation}
        \begin{pmatrix}
        a_s^{(o)} \\
        a_i^{\dagger(o)}
    \end{pmatrix} = \begin{pmatrix}
            \cosh(r) & e^{i\gamma}\sinh(r) \\
            e^{-i\gamma}\sinh(r) & \cosh(r) \\
        \end{pmatrix}
         \begin{pmatrix}
        a_s^{(i)} \\
        a_i^{\dagger(i)}
    \end{pmatrix}.
    \end{equation}
    In this equation, $a_s^{(o,i)}$ and $a_i^{(o,i)}$ are the signal and idler annihilation operators, where the superscript $(o)$ or $(i)$ denotes output or input modes, respectively. $r$ is a real number - the squeezing parameter - in the case of a second-order nonlinear interaction (e.g., down-conversion) in the PA. In the case of a third-order nonlinearity (e.g., four-wave mixing), $r$ has to be taken complex \cite{Ferrini2014}. $\gamma$ depends on the phase of the pump that drives the PA. Here, for simplicity, we will consider all PAs to be operated at $\gamma = 0$. This choice does not affect the generality of our derivation, since the pump phase serves merely as a reference for the phases of the signal and idler modes. We also define the shorthand notations: $G_k \equiv \cosh(r_k)$ and $g_k \equiv \sinh(r_k)$, where the index $k = 1,2$ corresponds to the relevant PA. For the SU(1,1) interferometer of Fig.\,\ref{Fig01}(a) exhibiting a phase shift $\phi$ between its two arms, the interferometer transformation (Appendix \ref{transformations}) is given as
    \begin{equation}\label{SU(1,1)_transformation}
        \begin{pmatrix}
        a_s^{(o)} \\
        a_i^{\dagger(o)}
    \end{pmatrix} = \begin{pmatrix}
            A & B \\
            D^* & C^* \\
        \end{pmatrix}
         \begin{pmatrix}
        a_s^{(i)} \\
        a_i^{\dagger(i)}
    \end{pmatrix} \ ,
    \end{equation}
    where the matrix elements are given by
    \begin{eqnarray}
        A &=& G_1G_2e^{i \phi} + g_1g_2\ ,\label{Eq04} \\
        B &=& G_2g_1e^{i \phi} + G_1g_2\ ,\label{Eq05}  \\
        C &=& G_1G_2 + g_1g_2e^{-i \phi}\ ,\label{Eq06} \\
        D &=& G_2g_1 + G_1g_2e^{-i \phi}. \label{Eq07}
    \end{eqnarray}\label{trnas coeff}
    One may note that $|A| = |C|$ and $|B| = |D|$ and $|A|^2 - |B|^2 = |C|^2 - |D|^2 = 1$. We now apply this transformation to obtain the expressions of the quadratures of the fields at the output ports, as a function of the relevant input state statistics.
    
\subsection{Evolution of the field quadratures}
    The field quadrature operator for the signal mode and reference phase $\theta$ is defined as
    \begin{equation}
        Y_s(\theta) = \frac{e^{i \theta}a_s - e^{-i \theta}a_s^{\dagger}}{i}\ ,
    \end{equation}
    with a similar expression for the idler mode. Using the interferometer transformation Eq.\,(\ref{SU(1,1)_transformation}), the quadrature observable at the output ports can be expressed as (see Appendix \ref{evolution} for a detailed calculation)
    \begin{align}
    Y_s^{(o)}(\theta)
    &= |A|Y_s^{(i)}(\theta + \kappa_{A})
       - |B|Y_i^{(i)}(\theta - \kappa_{B}), \\
    Y_i^{(o)}(\theta)
    &= |C|Y_i^{(i)}(\theta + \kappa_C)
       - |D|Y_s^{(i)}(\theta - \kappa_D),
    \end{align}
    where $\kappa_J$ is the argument of the complex coefficient $J = A, B, C, D$. Note that $\kappa_C = \kappa_A - \phi$ and $\kappa_D = \kappa_B - \phi$. In the following, we consider only the field quadratures with $\theta = 0$ and $\pi/2$. Like the pump phase $\gamma$, the value of $\theta$ simply shifts the interferometer response versus $\phi$, without affecting its sensitivity. We  adopt the short hand notations $Y_{s,i}(0) \equiv Y_{s,i}$ and $Y_{s,i}(\pi/2) \equiv X_{s,i}$. We now use these expressions to calculate the expectation values, fluctuations, and the resulting phase sensitivity that can be retrieved from the field quadratures measured at the output ports of the interferometer.

\subsection{Phase shift sensitivity in terms of quadratures}
    To estimate the phase shift, one measures an output observable that depends on this phase shift, which reads  $\phi = \phi_0 + \delta\phi$, where $\phi_0$ is a known bias point and $\delta\phi$ is the unknown phase shift variation to be estimated. The interferometer is said to be operated at the bright fringe when $\phi_0 = 2n\pi$ (the second PA amplifies the signal and idler) and at the dark fringe when $\phi_0 = (2n+1)\pi$ (when the second PA deamplifies the signal and idler), where $n$ is an integer. In this work, the measurement observable $O(\phi)$ is either a single output port quadrature $Y_s = -{i}(a_s-a_s^{\dagger})$ or $Y_i = -{i}(a_i-a_i^{\dagger})$ or a joint quadrature defined as $Y_{\pm}=Y_s \pm Y_i$. These observables can be obtained by the homodyne measurement schemes schematized in Fig.\,\ref{Fig01}(b). A small phase shift variation $\delta\phi$ produces a change in the expectation value given by $\frac{\partial \langle O(\phi) \rangle}{\partial \phi} \, \delta\phi$, which constitutes the phase-shift dependent signal. The sensitivity $\Delta \phi$ is then defined as the minimum value of $\delta\phi$ for which the change in $\langle O(\phi) \rangle$ is equal to the noise $\Delta O(\phi)$
    \begin{equation}
        \Delta \phi = \frac{\Delta O(\phi)}{\left| \partial \langle O(\phi) \rangle/\partial \phi \right|}\ .
    \end{equation}
    Sensitivity enhancement can therefore be achieved by increasing the response of the signal $\langle O(\phi) \rangle$ to phase changes or by reducing the fluctuations $\Delta O(\phi)$ of the measured observable, or both.
    
\section{Generalized sensitivity expressions}
    \label{sen exp}
    
    In what follows, we express the phase-shift sensitivity for a general two-mode quantum optical input state $\rho_{s,i}$ in terms of its mean quadrature amplitude and fluctuations. 
    
    In the first subsection, we begin with the ideal (lossless) SU(1,1) interferometer, using the transformations defined in Eq.\,(\ref{SU(1,1)_transformation}). Unless otherwise stated, the operating point is set at $\phi_0=\pi$, corresponding to the dark fringe. This choice fixes the relevant phase parameters as $\kappa_A=(2n+1)\pi$ and $\kappa_B=2n\pi$  where the integer $n$ is determined by the PA gain ratios $\frac{G_1}{g_1}$ and $\frac{G_2}{g_2}$. Note that this also fixes the values of $\kappa_C$ and $\kappa_D$.
    
    In the following subsection, we incorporate both internal losses (inside the interferometer arms) and external losses (in the output ports) into the formalism, as shown in Fig.\,\ref{Fig01}(a). The sensitivity is evaluated for each configuration using single-port and joint homodyne detection (see Fig.\,\ref{Fig01}(b)). In all cases, the full derivations are provided in the  Appendices \ref{statistic} and  \ref{sensitivity}.

\subsection{Ideal interferometer}

\subsubsection{Single port homodyne measurement}
\label{singleport}
    We now present the general expression for the phase sensitivity under single-port quadrature measurement, based on the formalism developed in the previous section. The measurement configuration corresponding to this case is illustrated to the left of Fig.\,\ref{Fig01}(b). Due to symmetry, the sensitivity expressions at the two output ports are equivalent up to relabeling of the input modes. We present the result for the observable $Y_s$ at the first output port, with the corresponding expression for $Y_i$ included in Appendix \ref{sensitivity}. At the dark fringe, the phase sensitivity for single quadrature measurement at the first port is given by
    \begin{widetext}
    \begin{eqnarray}\label{sens_1}
        (\Delta \phi)_{Y_s^{(o)}} = \frac{\sqrt{|A|^2(\Delta Y_s^{(i)})^2 + |B|^2(\Delta Y_i^{(i)})^2 + 2|A||B|(\braket{Y_s^{(i)}Y_i^{(i)}} - \braket{Y_s^{(i)}}\braket{Y_i^{(i)}})}}{G_2|(G_1\braket{X_s^{(i)} } + g_1\braket{X_i^{(i)}})|}.
    \end{eqnarray}
    \end{widetext}
    Note that the covariance term $(\braket{Y_s^{(i)}Y_i^{(i)}} - \braket{Y_s^{(i)}}\braket{Y_i^{(i)}})$ vanishes if the probe state is separable, i.e., if the input signal and the idler are not entangled. 
    Traditionally, the SU(1,1) interferometer is run in one of the two following PA gain-regimes: i) balanced gains ($G_1 = G_2 \equiv G$, $g_1 = g_2 \equiv g$) or ii) unbalanced gains ($G_2 \gg G_1$), where the gain of the second PA is increased to try to compensate for the losses that follow the first PA. This leads to different simplified versions of Eq.\,(\ref{sens_1}), which are given below. 
    
    \begin{center}{\textit{a. Balanced gains}}\end{center}
    
    In the case of balanced gains, at the dark fringe, Eqs.\,(\ref{Eq04}) and (\ref{Eq05}) lead to $|A|^2 = 1$ and $|B|^2 = 0$. The  sensitivity of Eq.\,(\ref{sens_1}) reduces to
    \begin{equation}
       (\Delta \phi)_{Y_s^{(o)}} = \frac{\Delta Y_s^{(i)}}{G|G\braket{X_s^{(i)} } + g\braket{X_i^{(i)}}|}. \label{Eq13}
    \end{equation}
    
    \begin{center}{\textit{b. Unbalanced gains}}\end{center}
    
    In the case of unbalanced gains, at the dark fringe, $|A|^2 \approx |B|^2 \approx G_2^2(G_1-g_1)^2$. The  sensitivity is then given by
    \begin{widetext}
    \begin{equation}
        (\Delta \phi)_{Y_s^{(o)}} =\frac{\sqrt{(\Delta Y_s^{(i)})^2 +(\Delta Y_i^{(i)})^2 + 2(\braket{Y_s^{(i)}Y_i^{(i)}} - \braket{Y_s^{(i)}}\braket{Y_i^{(i)}})}}{(G_1 + g_1)|G_1\braket{X_s^{(i)} } + g_1\braket{X_i^{(i)}}|}. \label{Eq14}
    \end{equation}  
    \end{widetext}

    \begin{center}{\textit{c. Discussion}}\end{center}

    These expressions (\ref{Eq13}) and (\ref{Eq14}) clarify the comparative strengths of the balanced and unbalanced gain operating regimes. Under balanced gain operation, at the dark fringe, Eq.\,(\ref{Eq13}) shows that the sensitivity becomes independent of $\Delta Y_i^{(i)}$, allowing enhancement through squeezing of the input signal mode, i.e., $\Delta Y_s^{(i)} \propto e^{-R}$ for a squeezing factor $R$. Additionally, increasing the average value of the input idler quadrature $\braket{X_i^{(i)}}$, improves the sensitivity without affecting the noise, since it does not contribute to output fluctuations. These strategies can be combined for optimal performance. This observation aligns with the findings of Li et al. \cite{23_li2014phase}, who identified the balanced gain regime as optimal for coherent and squeezed vacuum inputs. However, while their work established this result, it did not analyse sensitivity degradation under unbalanced operation. Our framework provides further insight into this issue.
    
    In the unbalanced gain regime, Eq.\,(\ref{Eq14}) shows that both strategies (reducing noise in only one input or increasing signal in the other input without impact on the noise) lose effectiveness. Since both input modes contribute equally to the output noise, squeezing a single input mode indeed fails to suppress the noise entering from the other input port. That said, when both input modes exhibit the same fluctuations, such as $\ket{\alpha}_s\otimes\ket{0}_{i}$, an unbalanced gain configuration provides better sensitivity. This agrees with prior results \cite{21_ou2020quantum}, though we emphasize that this advantage is limited to specific input states and offers only a constant-factor improvement in sensitivity, not a scaling advantage.
    
    Finally, although both configurations yield the same $1/G^2$ scaling in sensitivity, the mechanisms underlying this dependence differ. Although in both the single-port balanced case and the single-port unbalanced case, the phase-change signal scales in the same way ( Appendix \ref{sensitivity}, Eq. \ref{phasechangesignal}), in the case of balanced gains, the output variance  $\Delta Y_{s}^{(o)} = \Delta Y_{s}^{(i)}$ (Appendix \ref{statistic}, Eq. \ref{outvar} with $|A|^2 = 1$ and $|B|^2 = 0$ ).  On the contrary,  the unbalanced gain case yields $\Delta Y_{s}^{(o)} = G_2(G_1 - g_1)\sqrt{(\Delta (Y_{s}^{(i)} + Y_{i}^{(i)}))^2}$ (Appendix \ref{statistic}, Eq. \ref{outvar} with $|A|^2 \approx |B|^2 \approx G_2^2(G_1-g_1)^2$  ). These different sensitivity-enhancement mechanisms have a profound effect on the loss-tolerance properties of these two operating regimes, as we will see in Section \ref{Lossy}. Thus, selecting the optimal operating point depends on the input probe state and its statistics. A more detailed explanation of these two operating points is provided in Section \ref{Application} for the case of a $\ket{\alpha,0}_{s,i}$ input state.
    \\

\subsubsection{Dual-port/joint homodyne detection}
\label{joint}
    Similar to the single-port measurement case, in this section, we only give the final expression for the sensitivity in the case of joint homodyne measurement, relegating the detailed calculation to Appendix \ref{sensitivity}. This expression, valid both at dark and bright fringes for an arbitrary input state $\rho_{s,i}$, reads
    \begin{widetext}
    \begin{eqnarray}\label{sens_j}
        (\Delta \phi)_{Y_{\pm}^{(o)}} = \frac{\sqrt{|E|^2(\Delta Y_s^{(i)})^2 + |F|^2(\Delta Y_i^{(i)})^2 + 2|E||F|(\braket{Y_s^{(i)}Y_i^{(i)}} - \braket{Y_s^{(i)}}\braket{Y_i^{(i)}})}}{(G_2 \mp g_2)|(G_1\braket{X_s^{(i)}} + g_1\braket{X_i^{(i)}})|}\ ,
    \end{eqnarray}
    \end{widetext}
    where $E = A \mp D^*$ and $F = C \mp B^*$. The $\pm$ signs in $(\Delta \phi)_{Y_{\pm}^{(o)}}$ in Eq.\,(\ref{sens_j}) and in subsequent expressions throughout this article, correspond to the cases where the signals provided by the two homodyne detections are added or subtracted, respectively, as shown in the right panel of Fig.\,\ref{Fig01}(b).  At the dark fringe, one has $|E|^2 = |F|^2 = (G_1 \pm g_1)^2(G_2 \mp g_2)^2$. The expression for sensitivity at the dark fringe operating point then simplifies to
    \begin{widetext}
    \begin{equation}\label{joint dark}
       (\Delta \phi)_{Y_{\pm}^{(o)}} =  \frac{\sqrt{(\Delta Y_s^{(i)})^2 +(\Delta Y_i^{(i)})^2 + 2(\braket{Y_s^{(i)}Y_i^{(i)}} - \braket{Y_s^{(i)}}\braket{Y_i^{(i)}})}}{(G_1 \mp g_1)|(G_1\braket{X_s^{(i)} } + g_1\braket{X_i^{(i)}})|}\ . 
    \end{equation} 
    \end{widetext}
    Notice that this sensitivity does not depend on the second PA gain parameters $G_2$ and $g_2$. $Y_{-}^{(o)}$ would be here the operator to use at the dark fringe to minimize the sensitivity. In the case of bright fringe operation, i.e., $\phi_0 = 2n\pi$, $|E|^2 = |F|^2= (G_1 \mp g_1)^2(G_2 \mp g_2)^2$ and Eq.\,(\ref{sens_j}) becomes
    \begin{widetext}
    \begin{eqnarray}\label{joint bright}
        (\Delta \phi)_{Y_{\pm}^{(o)}} =\frac{\sqrt{(\Delta Y_s^{(i)})^2 +(\Delta Y_i^{(i)})^2 + 2(\braket{Y_s^{(i)}Y_i^{(i)}} - \braket{Y_s^{(i)}}\braket{Y_i^{(i)}})}}{(G_1 \pm g_1)|(G_1\braket{X_s^{(i)} } + g_1\braket{X_i^{(i)}})|}. \qquad
    \end{eqnarray} 
    \end{widetext}
    We notice that  $Y_{+}^{(o)}$ is the appropriate measurement operator at the bright fringe. As seen from Eq.\eqref{joint bright}, the noise in the  $Y_{+}^{(o)}$ quadrature is reduced compared to the combined noises of the two modes in their input state, showing that $Y_{+}^{(o)}$ is a squeezed quadrature. This noise suppression is precisely the mechanism by which the phase sensitivity is improved at the bright fringe. Although the gain in the phase-dependent signal is amplified by a relatively weak gain, the quadrature noise is squeezed enough so that the overall sensitivity becomes optimal (Appendix \ref{Sensitivity enhancement}). 
    
    In contrast, Eq.\eqref{joint dark} shows that $Y_{-}^{(o)}$ is the appropriate choice of measurement observable at the dark fringe. Indeed,  Eq.\eqref{joint dark} shows that the noise in  $Y_{-}^{(o)}$ cannot be reduced below the noise of the input states. At best, it can only match the input noise when the two PAs have the same gain. Nevertheless, the phase-dependent signal is significantly stronger at the dark fringe. This enhanced signal strength compensates for the lack of noise reduction, which is why the sensitivity improves at this operating point (Appendix \ref{Sensitivity enhancement}). With these optimal choices, i.e., $Y_{+}^{(o)}$ at the bright fringe and $Y_{-}^{(o)}$ at the dark fringe, the minimum achievable sensitivities from joint homodyne detection become identical to the unbalanced single-port homodyne detection. This result has already been reported in the literature \cite{27_anderson2017optimal,28_anderson2017phase} for the particular case of an input state $\ket{\alpha,0}_{s,i}$. However,  here we generalise this result to the case of an arbitrary choice of the input probe state $\rho_{s,i}$. 
    
    One important conclusion of this section is that in the case of an ideal, i.e. lossless, interferometer, the joint quadrature measurement and the unbalanced gain single-port measurement schemes yield the same sensitivity, as given in Eqs.\,(\ref{Eq14},\ref{joint dark},\ref{joint bright}). All the corresponding new expressions derived in this work for arbitrary input states are gathered in a formulary (Table \ref{table1}) of Appendix \ref{sen table}. Table \ref{table1} also recalls results from the existing literature obtained in the particular case of an injected coherent state in the input signal port that will be discussed in detail in Section \ref{Application}.
    \\
\subsection{Lossy interferometer}\label{Lossy}
All interferometric setups are subject to optical losses, both within and outside the interferometer, which reduce the sensitivity of the corresponding measurement. These photon losses can arise from various sources, including scattering, attenuation in optical fibers, coupling losses in waveguides, and imperfect detector efficiency. Both internal and external losses can respectively be modeled as beam splitters with an intensity transmission of $ 1-L_{s,i} $ or $ 1-l_{s,i} $ (see Fig.\,\ref{Fig01}(a)), respectively, within the interferometer arms or at the output ports.

\subsubsection{Internal losses}
    We consider only the two beam splitters placed in each arm between the first and the second PA in Fig.\,\ref{Fig01}(a), i.e., $L_s\neq 0$, $L_i\neq 0$, and $l_s=l_i= 0$. The single-port sensitivity at the dark fringe then becomes
    \begin{widetext}
    \begin{align}
    (\Delta \phi)_{Y_{s}^{(o)}} =
    \frac{
    \sqrt{
    |A_L|^2(\Delta Y_s^{(i)})^2
    + |B_L|^2(\Delta Y_i^{(i)})^2
    + 2|A_L| |B_L|
    \left(
    \braket{Y_s^{(i)}Y_i^{(i)}}
    - \braket{Y_s^{(i)}}\braket{Y_i^{(i)}}
    \right)
    + |A_L'|^2
    + |B_L'|^2
    }
    }{
    \sqrt{1-L_s}\,G_2\,
    \left|
    G_1\braket{X_s^{(i)}} + g_1\braket{X_i^{(i)}}
    \right|
    }.
    \label{noisy_sens_1}
    \end{align}
    \end{widetext}
    The expressions of the coefficients $A_L,B_L,A_L', B_L'$ are given in Appendix \ref{transformations}. Note that for the case of symmetric losses, i.e., $L_s = L_i \equiv L$, we have $A_L = A\sqrt{1-L}$ and $B_L = B\sqrt{1-L}$. For simplicity, we consider only this situation in the rest of the section. 

    \begin{center}{\textit{a. Balanced gains}}\end{center}
    Under this hypothesis and with the balanced gain condition $G_1=G_2\equiv G$ and $g_1=g_2\equiv g$, Eq.\,(\ref{noisy_sens_1}) reduces to
    \begin{equation}
        (\Delta \phi)_{Y_{s}^{(o)}} =\frac{\sqrt{(\Delta Y_s^{(i)})^2 + \eta}}{G|(G\braket{X_s^{(i)}} + g\braket{X_i^{(i)}})|},\label{noisy_1}
    \end{equation}
    in which, by comparison with Eq.\,(\ref{Eq13}),
    \begin{equation}
        \eta = \frac{L}{1-L}(G^2 + g^2)
        \label{eq19}
    \end{equation}
     can be interpreted as a correction factor with respect to the ideal case due to the presence of losses. 
    \\
 
    \begin{center}{\textit{b. Unbalanced gains}}\end{center}
    
    In the case of unbalanced gains ($G_2\gg G_1$), the presence of symmetric losses modifies Eq.\,(\ref{Eq14}) in the same manner, leading to 
    \begin{widetext}
    \begin{eqnarray}\label{noisy_2}
        (\Delta \phi)_{Y_{s}^{(o)}} =\frac{\sqrt{(\Delta Y_s^{(i)})^2 +(\Delta Y_i^{(i)})^2 + 2(\braket{Y_s^{(i)}Y_i^{(i)}} - \braket{Y_s^{(i)}}\braket{Y_i^{(i)}}) + \eta}}{(G_1 + g_1)|(G_1\braket{X_s^{(i)} } + g_1\braket{X_i^{(i)}})|}\ , 
    \end{eqnarray} 
    \end{widetext}
    with the correction factor $\eta$ with respect to Eq.\,(\ref{Eq14}) given by
    \begin{equation}
    \eta = \frac{2L}{1-L}(G_1 + g_1)^2\ .
    \label{eq21}
    \end{equation}
    
    \vspace{5pt}
    \begin{center}{\textit{c. Dual port/joint homodyne detection}}\end{center}
    
    In the case of dual-port joint homodyne measurement at both dark and bright fringes, the presence of losses modifies Eq.\,(\ref{sens_j}) to
    \begin{widetext}
    \begin{align}
    (\Delta \phi)_{Y_{\pm}^{(o)}} =
    \frac{
    \sqrt{
    |E_L|^2(\Delta Y_s^{(i)})^2
    + |F_L|^2(\Delta Y_i^{(i)})^2
    + 2|E_L||F_L|
    \left(
    \braket{Y_s^{(i)}Y_i^{(i)}}
    - \braket{Y_s^{(i)}}\braket{Y_i^{(i)}}
    \right)
    + |E_L'|^2
    + |F_L'|^2
    }
    }{
    \sqrt{1-L_s}\,(G_2 \mp g_2)\,
    \left|
    G_1\braket{X_s^{(i)}} + g_1\braket{X_i^{(i)}}
    \right|
    }\ .
    \label{noisy_sens_j}
    \end{align}
    \end{widetext}
    
    The expressions for the coefficients $E_L, F_L, E_L', F_L'$ are given in  Appendix \ref{evolution}. Again, for the case of symmetric losses, we have $E_L = E\sqrt{1-L}$ and $F_L = F\sqrt{1-L}$. At the dark fringe, the resulting expression for the sensitivity becomes
    \begin{widetext}
    \begin{eqnarray}\label{noisy_3}
        (\Delta \phi)_{Y_{\pm}^{(o)}} =\frac{\sqrt{(\Delta Y_s^{(i)})^2 +(\Delta Y_i^{(i)})^2 + 2(\braket{Y_s^{(i)}Y_i^{(i)}} - \braket{Y_s^{(i)}}\braket{Y_i^{(i)}}) + \eta}}{(G_1 \mp g_1)|(G_1\braket{X_s^{(i)} } + g_1\braket{X_i^{(i)}})|}\ , 
    \end{eqnarray} 
    \end{widetext}
    with 
    \begin{equation}
        \eta = \frac{2L}{1-L}(G_1 \mp g_1)^2\ .
    \end{equation}
    Similarly, the expression for the sensitivity at the bright fringe is modified by the presence of the losses to become
    \begin{widetext}
    \begin{equation} \label{noisy_4}
        (\Delta \phi)_{Y_{\pm}^{(o)}} =\frac{\sqrt{(\Delta Y_s^{(i)})^2 +(\Delta Y_i^{(i)})^2 + 2(\braket{Y_s^{(i)}Y_i^{(i)}} - \braket{Y_s^{(i)}}\braket{Y_i^{(i)}}) + \eta}}{(G_1 \pm g_1)|(G_1\braket{X_s^{(i)} } + g_1\braket{X_i^{(i)}})|}\ ,
    \end{equation} 
    \end{widetext}
    where
    \begin{equation}
    \eta = \frac{2L}{1-L}(G_1 \pm g_1)^2\ .
    \end{equation}
    We note that even in the presence of internal losses, the sensitivities of the joint measurements interchange roles: the sensitivity of $Y_{+}^{(o)}$ at the bright fringe matches that of $Y_{-}^{(o)}$ at the dark fringe, and vice versa, as it was in the ideal scenario discussed in Section \ref{joint}. With respect to the lossless case, both measurement sensitivities are affected by the same correction factor $\eta = \frac{2L}{1-L}(G_1 + g_1)^2$. A complete summary of all generalized sensitivity expressions in the presence of internal losses is presented in the Appendix \ref{sen table} (Table \ref{table2}) for ease of comparison.
    
\subsubsection{External losses}
\label{external losses}
Previous studies, though mainly restricted to number measurements, reported that SU(1,1) interferometers are more resistant to external losses compared to internal losses \cite{30_hudelist2014quantum,31_marino2012effect,32_ou2012enhancement}. This is because internal losses couple vacuum noise to the correlated modes inside the interferometer, and this noise is amplified by the second PA, thereby increasing its magnitude. On the other hand, the noise due to external losses located just before the detection is not amplified and remains small compared to the highly amplified signal at the output of the interferometer. Here, we study the impact of external losses for homodyne measurements and observe that by adjusting the gain of the second amplifier,  it is possible to compensate for external losses only for a few specific homodyne measurements, as detailed below.  

Let us thus consider the two beam splitters of a reflectivity $l_k$ placed in each arm between the second PA and the measurement devices (see Fig.\,\ref{Fig01}a), leading to $l_k\neq0$, but $L_k=0$ (zero internal losses) for $k = s,i$. Let us first consider the case where only the output signal mode is detected. Then, with $A_l = \sqrt{1-l_s}A$ and $B_l = \sqrt{1-l_s}B$, the expression for the sensitivity at the dark fringe becomes
\begin{widetext}
\begin{eqnarray}\label{external1}
    (\Delta \phi)_{Y_{s}^{(o)}} = \frac{\sqrt{|A_l|^2(\Delta Y_s^{(i)})^2 + |B_l|^2(\Delta Y_i^{(i)})^2 + 2|A_l||B_l|(\braket{Y_s^{(i)}Y_i^{(i)}} - \braket{Y_s^{(i)}}\braket{Y_i^{(i)}}) + l_s}}{\sqrt{1-l_s}G_2|(G_1\braket{X_s^{(i)}} + g_1\braket{X_i^{(i)}})|}.
\end{eqnarray}
\end{widetext}
We note that the loss parameter $l_i$ does not affect the sensitivity when only the signal is detected. The sensitivity, in the balanced gain condition ($G_1=G_2$ and $g_1=g_2$) and at the dark fringe is given by Eq.\,(\ref{noisy_1}),  but with the correction factor given as
\begin{equation}\label{corr1}
 \eta = \frac{l_s}{1-l_s}   
\end{equation} 
Notably, the correction factor in Eq.\,\eqref{corr1} is independent of the PA gain, whereas for the internal losses the correction factor (Eq.\,\eqref{eq19}) is gain dependent. 

Similarly, the sensitivity at the dark fringe for unbalanced gain operation remains the same as in Eq.\,(\ref{noisy_2}), provided one replaces the correction factor by 
\begin{equation}\label{corr2}   
\eta = \frac{l_s}{1-l_s}\left(\frac{G_1+g_1}{G_2}\right)^2.
\end{equation}
It should be noted that the correction factor in Eq.\,\eqref{corr2} depends on the gain of the second PA, which was not the case for internal losses (Eq.\,\eqref{eq21}). Therefore, the effect of external losses can be mitigated by increasing the gain of the second amplifier, as previously verified for number measurements \cite{30_hudelist2014quantum,31_marino2012effect,32_ou2012enhancement}. This explains why, in the unbalanced-gain configuration of the SU(1,1) interferometer, it is generally preferable for the second PA to have a higher gain than the first, as adopted in this study. 

Finally, we derive the expression for the sensitivity in the case of joint homodyne measurements. At the dark fringe, the sensitivity becomes
\begin{widetext}
\begin{eqnarray}\label{external2}
    (\Delta \phi)_{Y_{\pm}^{(o)}} = \frac{\sqrt{|E_l|^2(\Delta Y_s^{(i)})^2 + |F_l|^2(\Delta Y_i^{(i)})^2 + 2|E_l||F_l|(\braket{Y_s^{(i)}Y_i^{(i)}} - \braket{Y_s^{(i)}}\braket{Y_i^{(i)}}) + l_s + l_i}}{(\sqrt{1-l_s}G_2 \mp \sqrt{1-l_i}g_2)|(G_1\braket{X_s^{(i)}} + g_1\braket{X_i^{(i)}})|} \ ,
\end{eqnarray}
\end{widetext}
where the explicit definitions of the coefficients $E_l,F_l$ are given in Appendix \ref{evolution}. For the case of symmetric losses $l\equiv l_s=l_i$, the sensitivity at the dark fringe in the case of  joint signal and idler measurement is given by Eq.\,\eqref{noisy_3}, but with a correction factor given as
\begin{equation}\label{corr3}
\eta = \frac{2l}{1-l}\left(\frac{G_2 \pm g_2}{G_1 \pm g_1}\right)^2\ .     
\end{equation}

On the contrary, for the bright fringe operation, we observe that the sensitivity is given by Eq.\,(\ref{noisy_4}) with a correction factor
\begin{equation}\label{corr4}
\eta = \frac{2l}{1-l}\left(\frac{G_2 \pm g_2}{G_1 \mp g_1}\right)^2\ .     
\end{equation}
In the case of joint-homodyne measurement, an asymmetry arises between the two operating conditions. At the dark fringe, the correction factor becomes independent of the gains of the two PAs when the gains are equal; however, in the unbalanced-gain configuration ($G_2 \gg G_1$), it scales inversely with the square of the gain of the second PA. In contrast, at the bright fringe under the balanced-gain condition, the correction factor increases quartically with the PA gains, leading to much larger fluctuations for comparable levels of loss. Consequently, the impact of external losses can be reduced by increasing the gain of the second PA only in dark-fringe operation, whereas for bright-fringe operation, the situation is reversed. Nevertheless, in the presence of external losses, dark-fringe operation consistently outperforms bright-fringe operation for any choice of PA gains, although in the ideal lossless case, both configurations provide similar sensitivity. This stronger resilience to external losses in dark-fringe operation arises because the sensitivity enhancement is associated with a stronger phase-sensing signal. In contrast, at the bright-fringe operating point, the enhanced sensitivity originates from quadrature squeezing, which degrades rapidly in the presence of external losses.




All newly derived expressions for an SU(1,1) interferometer with external losses and arbitrary input states are summarized in the formulary (Table \ref{table3}) of Appendix \ref{sen table}, which highlights our findings. The table also includes the corresponding expressions for the specific case of an injected coherent state, which is discussed in detail in Section \ref{Application}. 
\\

\section{Truncated interferometer}
Let us consider now the setup shown in Fig.\,\ref{Fig01}(c), where the second PA of the SU(1,1) interferometer is removed, and the measurement is performed right after the phase shift to be measured. This setup is known as the truncated SU(1,1) interferometer \cite{28_anderson2017phase,29_gupta2018optimized}. We first note the following property of this truncated interferometer: it is somewhat obvious that since the two output beams of the first PA do not interfere with each other, one has to use the idler beam as a reference, in conjunction with the signal beam, to measure the phase shift. This is achieved in practice by measuring some quadratures of both output ports of the truncated interferometer and electronically adding or subtracting their signals, i.e., by performing a joint homodyne measurement.

\subsection{Ideal truncated interferometer}
The transformations for the ideal truncated interferometer are the same as the ideal full interferometer (see Eq.\,\ref{SU(1,1)_transformation}), but with different expressions of the coefficients $A,...,F$. In the present case, they become $A = G_1e^{i\phi}$, $B = g_1e^{i\phi}$, $C = G_1$, and $D =g_1$. Then, for $E$ and $F$ defined in the same way as the full interferometer, namely $E = A \mp D^*$ and $F = C \mp B^*$, the sensitivity of the  joint measurement at both bright and dark fringes becomes
\begin{widetext}
\begin{eqnarray}\label{Eq26}
    (\Delta \phi)_{Y_{\pm}^{(o)}} = \frac{\sqrt{|E|^2(\Delta Y_s^{(i)})^2 + |F|^2(\Delta Y_i^{(i)})^2 + 2|E||F|(\braket{Y_s^{(i)}Y_i^{(i)}} - \braket{Y_s^{(i)}}\braket{Y_i^{(i)}})}}{|(G\braket{X_s^{(i)}} + g\braket{X_i^{(i)}})|}\ ,
\end{eqnarray}
\end{widetext}
At the dark fringe, we have $|E|^2 = |F|^2 = (G_1 \pm g_1)^2$. The expression of the sensitivity then reduces to
\begin{widetext}
\begin{eqnarray}\label{truc bright}
    (\Delta \phi)_{Y_{\pm}^{(o)}} =\frac{\sqrt{(\Delta Y_s^{(i)})^2 +(\Delta Y_i^{(i)})^2 + 2(\braket{Y_s^{(i)}Y_i^{(i)}} - \braket{Y_s^{(i)}}\braket{Y_i^{(i)}})}}{(G_1 \mp g_1)|(G_1\braket{X_s^{(i)} } + g_1\braket{X_i^{(i)}})|}. 
\end{eqnarray} 
\end{widetext}
At the bright fringe, similarly, one has $|E|^2 = |F|^2 = (G_1 \mp g_1)^2$ and the resulting expression for the sensitivity becomes
\begin{widetext}
\begin{eqnarray}\label{truc dark}
    (\Delta \phi)_{Y_{\pm}^{(o)}} =\frac{\sqrt{(\Delta Y_s^{(i)})^2 +(\Delta Y_i^{(i)})^2 + 2(\braket{Y_s^{(i)}Y_i^{(i)}} - \braket{Y_s^{(i)}}\braket{Y_i^{(i)}})}}{(G_1 \pm g_1)|(G_1\braket{X_s^{(i)} } + g_1\braket{X_i^{(i)}})|}. 
\end{eqnarray} 
\end{widetext}
These expressions are identical to the ones for the full interferometer. Here again the best sensitivity is given by measuring respectively  $Y_{+}^{(o)}$ at the bright fringe and $Y_{-}^{(o)}$ at the dark fringe. This result has already been reported for the particular case of an input state $\ket{\alpha,0}_{s,i}$, but the preceding expressions generalize this result to any arbitrary input probe quantum state $\rho_{s,i}$. The output noise for both operators is lower than the combined noise of the two input modes. On the other hand, the phase-dependent signal is improved by the gain factor of the PA with respect to the input quadrature strength. So the sensitivity enhancement in this truncated configuration is achieved through both noise reduction and phase-dependent signal enhancement, both at bright fringe and dark fringe (Appendix \ref{Sensitivity enhancement}).

\subsection{Lossy truncated interferometer}
We model the losses of the truncated interferometer by introducing two beam splitters before the two homodyne detections with loss factors $l_k$ for $k = s,i$. The expression of the sensitivity in the presence of these losses then becomes
\begin{widetext}
\begin{eqnarray}\label{Eq29}
    (\Delta \phi)_{Y_{\pm}^{(o)}} = \frac{\sqrt{|E_l|^2(\Delta Y_s^{(i)})^2 + |F_l|^2(\Delta Y_i^{(i)})^2 + 2|E_l||F_l|(\braket{Y_s^{(i)}Y_i^{(i)}} - \braket{Y_s^{(i)}}\braket{Y_i^{(i)}}) + l_s + l_i}}{\sqrt{1-l_s}|(G_1\braket{X_s^{(i)}} + g_1\braket{X_i^{(i)}})|}.
\end{eqnarray}
\end{widetext}
The explicit form of the lossy interferometer transformations, along with a definition of the constants used above, is given in Appendix \ref{transformations}. In the case of symmetric losses, i.e., $l_s=l_i\equiv l$, sensitivity at the dark fringe is given by Eq. (\ref{noisy_3}) with the following correction factor:
\begin{equation}
\eta = \frac{2l}{1-l}(G_1 \mp g_1)^2\ .
\end{equation}
On the contrary, for the bright fringe operation, we observe that the sensitivity is given by Eq.\,(\ref{noisy_4}) with a correction factor:
\begin{equation}
\eta = \frac{2l}{1-l}(G_1 \pm g_1)^2\ .
\end{equation}
Notice that in this truncated configuration, the losses $l_s$ and $l_i$ are not really external losses, in the sense of the external losses of the configuration of Fig.\,\ref{Fig01}. Indeed, these losses are located before the combined homodyne detection, which in this configuration plays the role of the second beam splitter that recombines the signal and idler modes. This explains why in the truncated interferometer, these losses play a similar role to the internal losses of the full interferometer under joint measurement.

\section{Application: Example of a $\ket{\alpha,0}_{s,i}$ probe state}\label{Application}
The SU(1,1) interferometer with an input state $\ket{\alpha}_s \otimes \ket{0}_i$ has been widely explored in both theory \cite{21_ou2020quantum} and experiment, across platforms ranging from atomic ensembles to fiber-based systems \cite{13_lukens2018broadband,15_liu2019optimum,16_manceau2017detection,30_hudelist2014quantum}. As discussed in section \ref{Lossy}, both internal and external losses pose a major limitation to the ability of these interferometers to surpass the SQL \cite{31_marino2012effect,32_ou2012enhancement}. While the effect of losses on homodyne detection was previously addressed by Ou \cite{32_ou2012enhancement}, their treatment was limited to balanced configurations and did not account for a general noise analysis across different homodyne detection strategies. Our comprehensive operator-based framework fills this gap by enabling a complete and transparent comparison of different homodyne detection schemes, i.e., balanced, unbalanced single-port, and joint homodyne under both ideal and lossy conditions. While different schemes may achieve the same minimum sensitivity in the lossless limit, our analysis reveals that they rely on distinct physical mechanisms, as discussed in Section \ref{sen exp}. Balanced homodyne detection operates at the vacuum noise level, while unbalanced detection operates at a higher noise level under ideal conditions. However, the introduction of losses reintroduces vacuum noise, which disproportionately degrades the unbalanced scheme. Moreover, external losses modify the scaling behaviour of sensitivity for single-port unbalanced homodyne and joint homodyne cases (except with equal gain at the dark fringe operating point), contradicting earlier assumptions \cite{32_ou2012enhancement}.

In this final section, we apply our general formalism to perform a detailed sensitivity analysis for the $\ket{\alpha}_s \otimes \ket{0}_i$ input state. Although this configuration is well-established, our fully analytical formalism not only reproduces the known results but also provides new insights into the underlying noise-signal trade-off for different measurement strategies. This serves both as validation of our general approach and as a demonstration of the deeper physical understanding provided by our comprehensive analytical model. We systematically analyse sensitivity in the presence of internal and external losses, compare the performance of various homodyne detection schemes, and identify the precise conditions under which the SU(1,1) interferometer maintains an advantage over the classical MZI.

\subsection{Ideal interferometer}
We now apply the general framework developed in section \ref{sen exp} to $\ket{\alpha,0}_{s,i}$ input to the SU(1,1) interferometer. The analytical expressions for single-port balanced and unbalanced homodyne and joint homodyne are well known \cite{21_ou2020quantum}, and here we reproduce them directly within our formalism to confirm the consistency with established results. As the input state is separable, there is no correlation between the quadratures of the two modes, and the covariance among the quadratures is therefore zero. With the quadrature conventions defined in Section \ref{probe formalism}, each mode carries unit quadrature noise. Incorporating these conditions into the expressions obtained in Section \ref{sen exp}, we obtain the phase shift sensitivity for single-port homodyne detection at an arbitrary operating phase $\phi_0$ as
\begin{align}
(\Delta \phi)_{Y_{s}^{(o)}} =
\frac{
\sqrt{
(G_1^2+g_1^2)(G_2^2+g_2^2)+4G_1G_2g_1g_2\cos(\phi)
}
}{
2G_1G_2|\cos(\phi)|\,|\alpha|
}\,.
\label{noisy_sens_clean}
\end{align}

The  sensitivity for joint homodyne detection is likewise given by the following expression
\begin{eqnarray}
(\Delta\phi)_{Y_{\pm}^{(o)}} = \frac{\sqrt{G_1^2 \mp 2G_1g_1{\cos{\phi}} + g_1^2}}{\sqrt{2}G_1|\cos{(\phi)}||\alpha|}.
\end{eqnarray}
We recall from Section \ref{sen exp} that $Y_{+}^{(o)}$ is optimal at the bright fringe and $Y_{-}^{(o)}$ at the dark fringe. Note that Eq.\,(\ref{sens_1}) and Eq.\,(\ref{sens_j}) give sensitivity specifically at the dark fringe and bright fringes, respectively. The general sensitivity formula applicable to any operating phase is presented in Appendix \ref{sensitivity}.

In Fig.\,\ref{fig:epsart}, we  plot the sensitivity for all measurement schemes as a function of the phase shift bias $\phi_0 $. All observables except $Y_+$ attain their minimum sensitivity near the dark fringe at $\phi_0 = \pi$. Near the dark fringe, the unbalanced single-port homodyne and joint homodyne $(Y_-)$ schemes exhibit the same minimum sensitivity. The $Y_+$ observable achieves an equal minimum sensitivity at the bright fringe ($\phi_0 =0, 2\pi$). To make a comparison with a classical MZI with the same input coherent states, all sensitivity values are normalized by the square root of the input photon number of the coherent state $|\alpha|$, such that the sensitivity of a classical Mach–Zehnder interferometer under the same input field strength is unity. We observe that across a wide range of phase shifts, the SU(1,1) interferometer offers superior sensitivity compared to the classical MZI.

\begin{figure}[htp]
\centering
\includegraphics[width=0.45\textwidth]{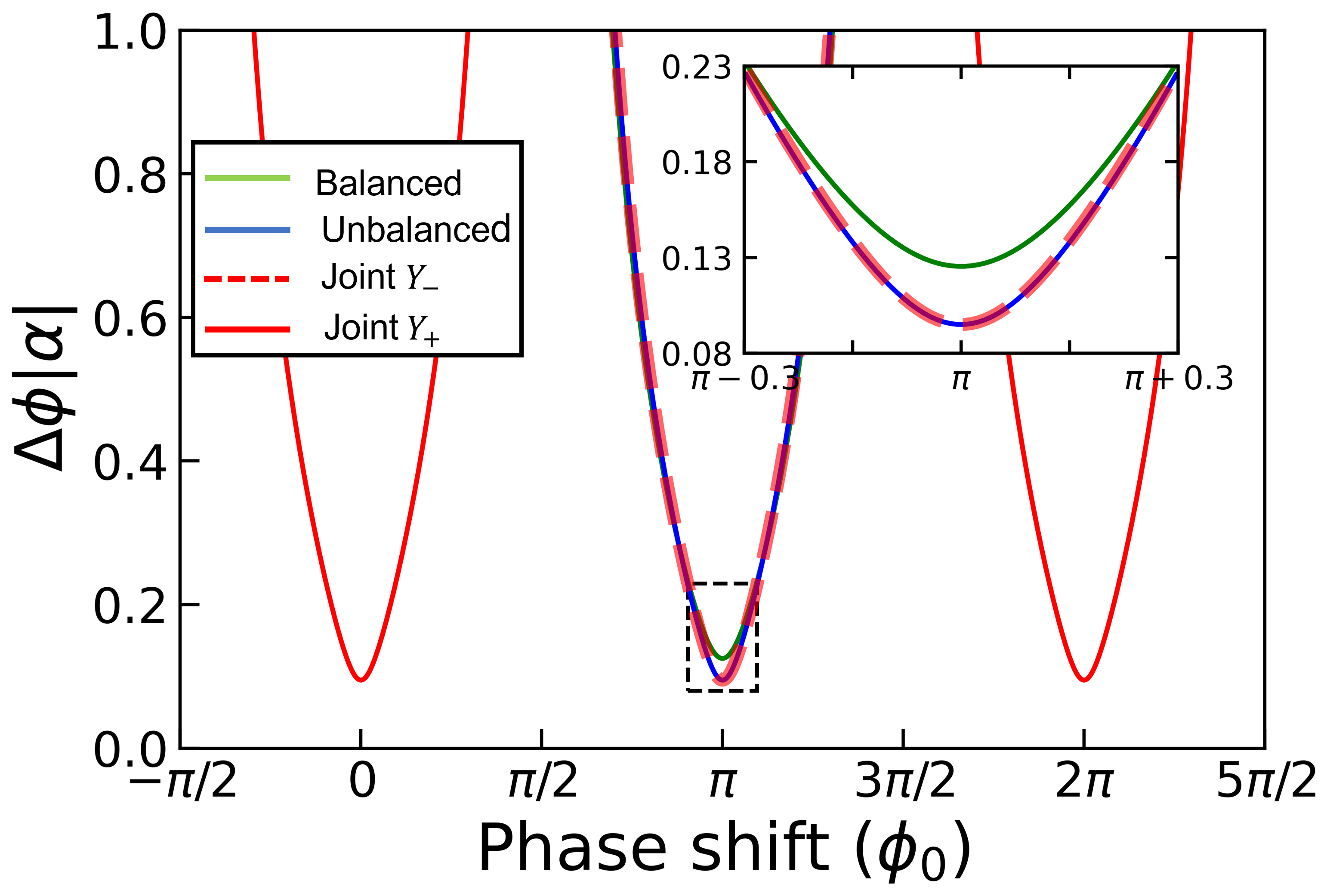} 
\caption{\label{fig:epsart} Phase sensitivity as a function of the phase shift bias $\phi_0=\phi-\delta\phi$. The phase of the coherent input with respect to the external reference is taken to be zero. The plots correspond to balanced and joint homodyne configuration with PA gains $G_1 = G_2 = 2$ and an unbalanced single-port homodyne case with $G_1=2$ and $G_2=5$. The vertical axis is calibrated with respect to the phase shift sensitivity of a Mach-Zehnder interferometer, which achieves $\Delta \phi|\alpha|=1$ } 
\end{figure}

In the case of balanced gains, the phase sensitivity reaches its minimum at $\phi_0 = \pi$, and is given by $\frac{1}{2G^2|\alpha|}$. At this point, the output noise equals the input noise while the signal is amplified; this effect is known as quantum noiseless amplification \cite{21_ou2020quantum}. For the unbalanced gain configuration, the sensitivity minimum occurs at the same operating point and is given by $\frac{1}{\sqrt{2}G_1(G_1+g_1)|\alpha|}$. It is clear that for this type of input, the unbalanced configuration has an improved sensitivity with respect to the balanced gain configuration by a factor of $\sqrt{2}$ in the high gain limit (see the inset of Fig.\,\ref{fig:epsart}). Numerically, we notice that for $G_1=2$, the high gain limit $G_2\gg G_1$ is approached approximately at $G_2=5$, which corresponds to the plots of Fig.\,\ref{fig:epsart}. Figure\,\ref{fig:epsart} also shows that the sensitivity for joint homodyne measurement reaches the same minimum value as the unbalanced gain single-output-port measurement. Although both $Y_{+}$ and $Y_{-}$ in a joint measurement yield the same minimum sensitivity, the underlying mechanisms differ significantly. At the dark fringe, the signal corresponding to $Y_-$ scales with $G^2$ while the noise remains the same as the input one. In contrast, at the bright fringe, the signal associated with $Y_+$ remains unity while the noise scales inversely with $G^2$.

The conclusion in the case of the ideal interferometer is that the joint homodyne detection and the unbalanced gain configurations only permit improving the sensitivity by a factor of $\sqrt{2}$ with respect to the case of the single-output-port homodyne detection with balanced gains.

\subsection{Lossy interferometer}

We now present the analytical expressions for phase sensitivity in the presence of internal losses, which significantly affect the scaling behaviour of the SU(1,1) interferometer and limit its ability to surpass the SQL. The full analytical expressions for phase sensitivity, including arbitrary losses in both arms, for single-port (balanced and unbalanced gains) and joint homodyne detection, are given by Eqs.\,(\ref{noisy_sens_1},\ref{noisy_sens_j}), respectively, in Section \ref{sen exp}, and are not particularly intuitive. However, when identical losses are assumed in both arms, the minimum sensitivity in all configurations occurs at the same operating point as in the ideal (lossless) case. In all scenarios, the effect of losses manifests itself as an additive correction to the expression of ideal sensitivity. Assuming identical losses in both arms, the minimum sensitivity for the balanced single-port homodyne case from Eq. (\ref{noisy_1}) is given by 

\begin{figure}[htp]
\centering
\includegraphics[width=\linewidth]{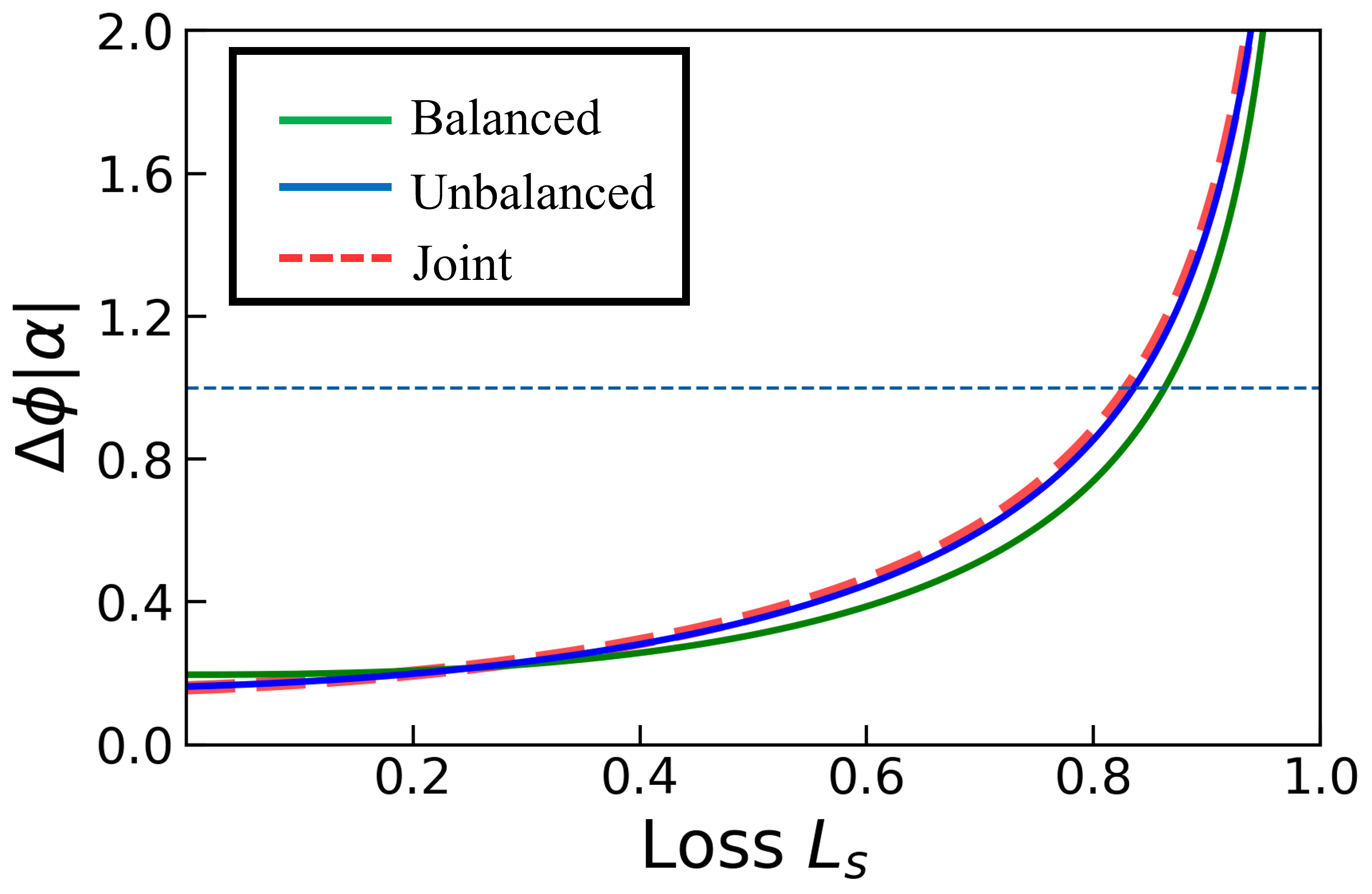} 
\vspace{2pt}
\textbf{(a)} 
\vspace{6pt} 
\includegraphics[width=\linewidth]{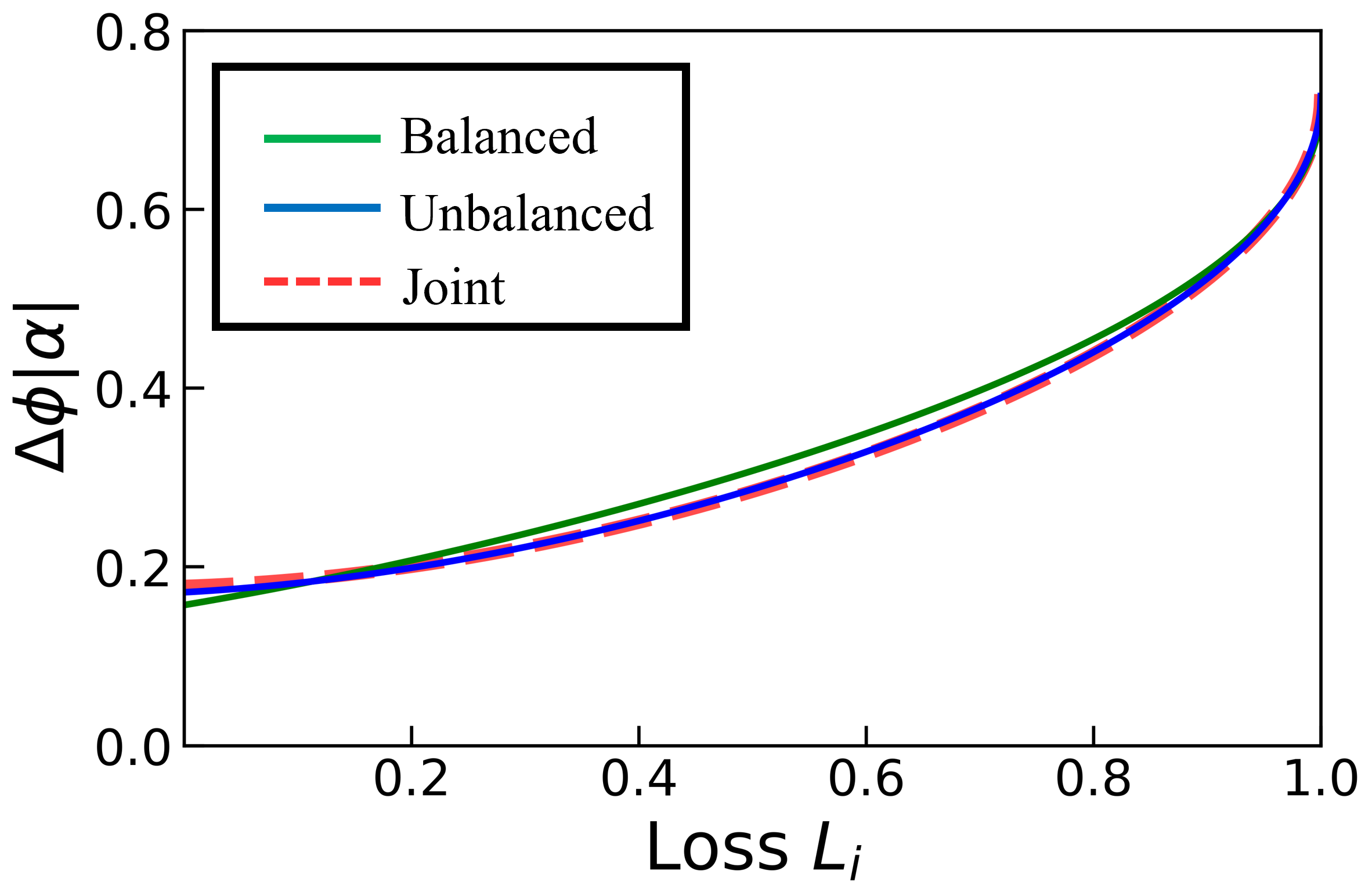} 
\vspace{2pt}
\textbf{(b)} 
\vspace{6pt} 

\caption{\label{Fig03} Phase sensitivity as a function of asymmetric internal loss for the three detection schemes. Obtained from Eqs.  \ref{noisy_sens_1} and \ref{noisy_sens_j}. Gain values used for these plots: balanced gain configuration: $G=2$; unbalanced gain configuration: $G_1=2,G_2=5$; joint homodyne detection: $G=2$. Together, these plots illustrate the asymmetric impact of losses in the two arms.  \textbf{(a)} The losses $L_s$ in the signal arm are varied while the losses in the other arm are kept fixed ($L_i=0.2$). The dotted blue horizontal line corresponds to the phase shift sensitivity of a Mach-Zehnder interferometer which achieves $\Delta \phi|\alpha|=1$ \textbf{(b)}. The losses $L_i$ in the idler arm are varied while the losses in the other arm remain fixed ($L_s=0.2$).}
\end{figure}

\begin{eqnarray} \label{28.sph}
    (\Delta \phi)_{Y_{s}^{(o)}} =\frac{\sqrt{1 + \frac{L}{1-L}(G^2+g^2)}}{2G^2|\alpha|}.
\end{eqnarray}
Moreover, for the case of unbalanced gains with single-port detection and the case of joint homodyne detection, we obtain,  from Eqs. (\ref{noisy_2}), (\ref{noisy_3}), and (\ref{noisy_4}) 
\begin{align}
(\Delta \phi)_{Y_{\pm}^{(o)}} = (\Delta \phi)_{Y_{s}^{(o)}} =
\frac{
\sqrt{
1+\frac{L}{1-L}(G_1 + g_1)^2
}
}{
\sqrt{2}G_1(G_1+g_1)|\alpha|
}\,.
\label{29.dhd}
\end{align}

One interesting observation is that when the losses in both arms are equal, the correction in sensitivity due to internal losses is the same for the schemes with single-port detection with unbalanced gains and joint homodyne detection with equal gains. It is evident from Eqs. (\ref{28.sph}) and (\ref{29.dhd}) that, for a given level of internal losses and fixed PA gain, the additive correction term is larger in the unbalanced gain and joint homodyne detection cases, leading to degradation of sensitivity relative to the balanced configuration. At high parametric gain, the influence of the losses becomes increasingly significant, altering the scaling behaviour of the phase sensitivity. In the ideal case without losses, the sensitivity improves as $1/G^2$, however, in the presence of losses, the scaling degrades to $1/G$ asymptotically. Thus, while increasing gain still improves sensitivity, the improvement decreases progressively at large values of $G$.

\begin{figure}[htp]
\centering
\includegraphics[width=\linewidth]{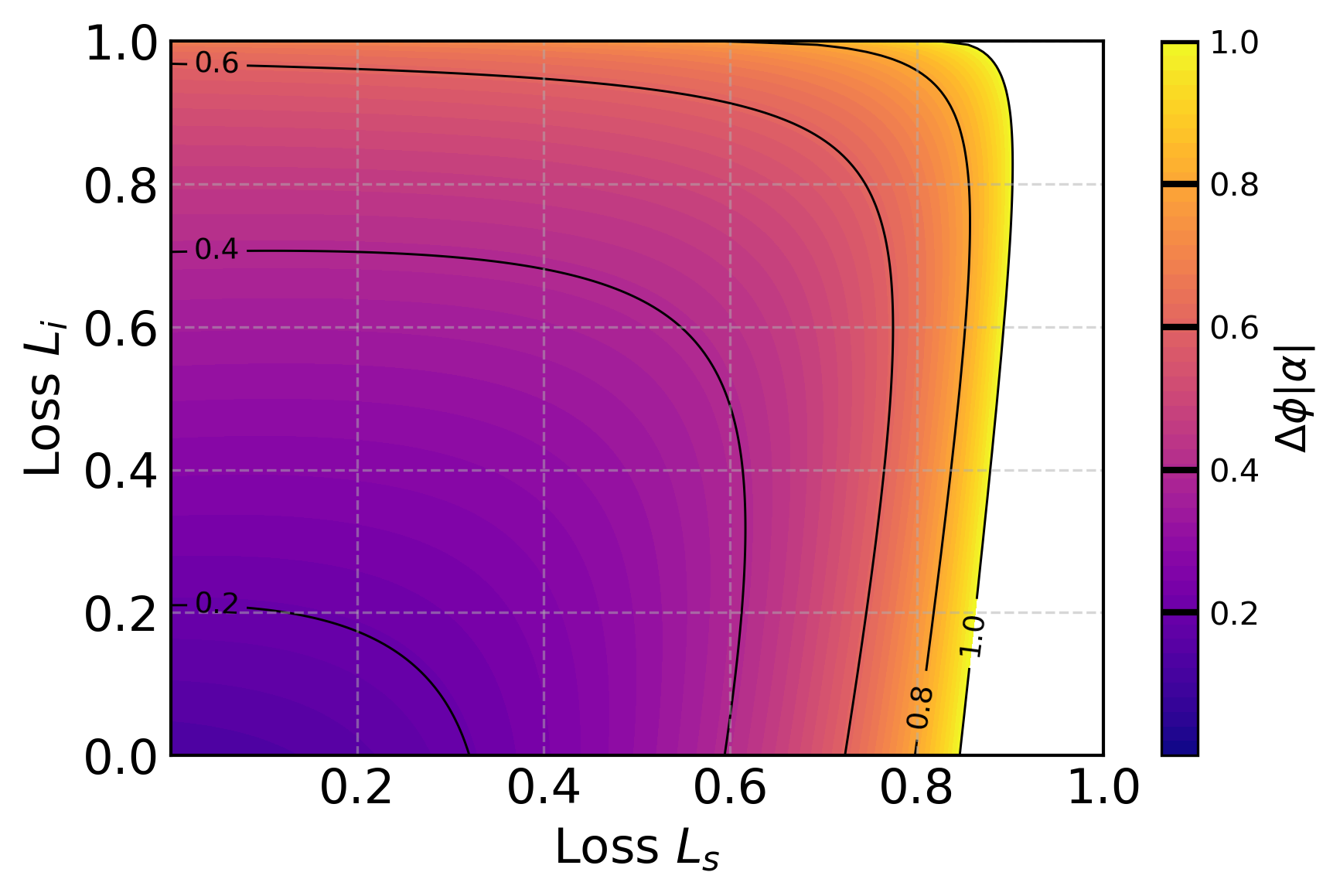} 
\vspace{2pt}
\textbf{(a)} 
\vspace{6pt} 
\includegraphics[width=\linewidth]{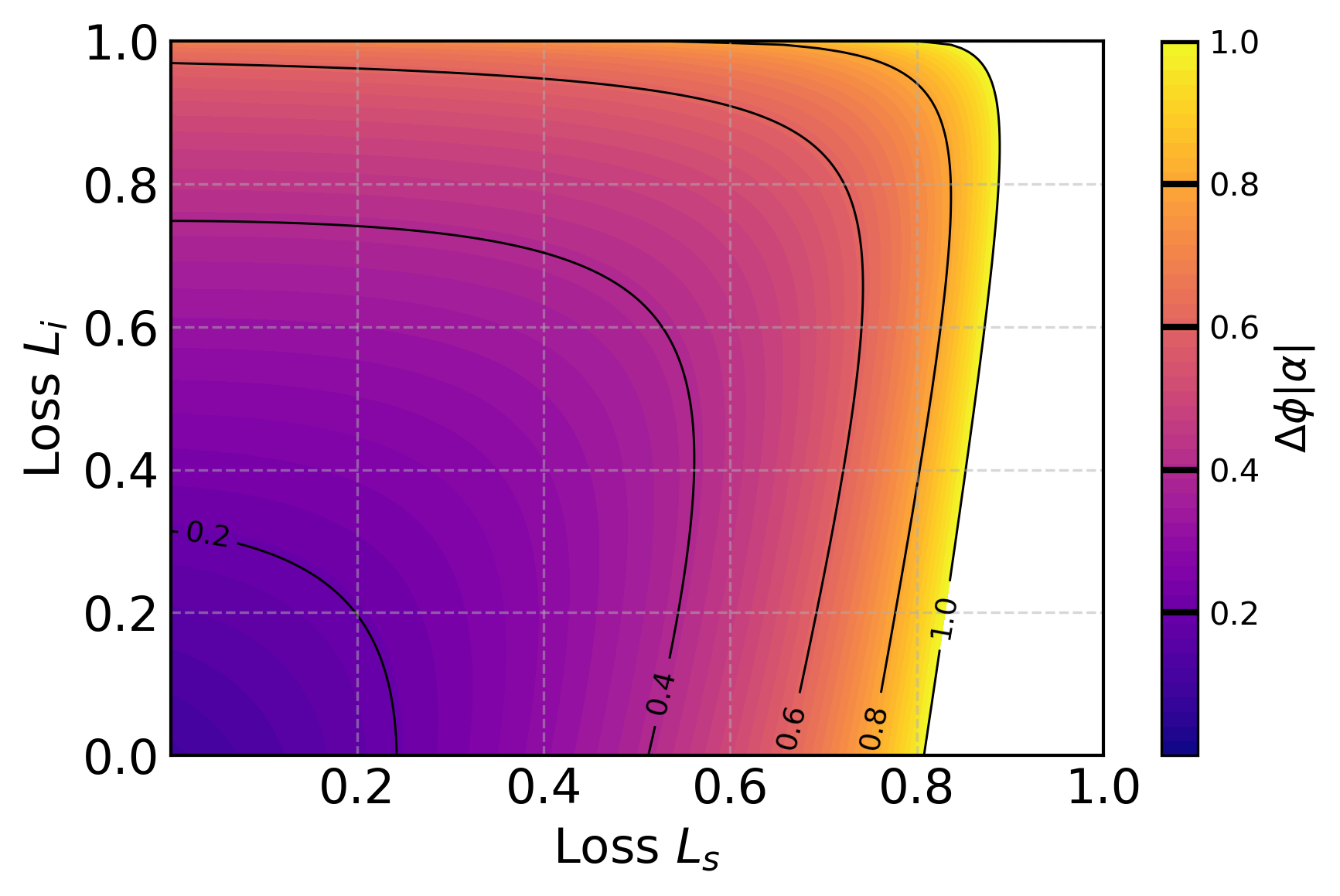} 
\vspace{2pt}
\textbf{(b)} 
\vspace{6pt}
\includegraphics[width=\linewidth]{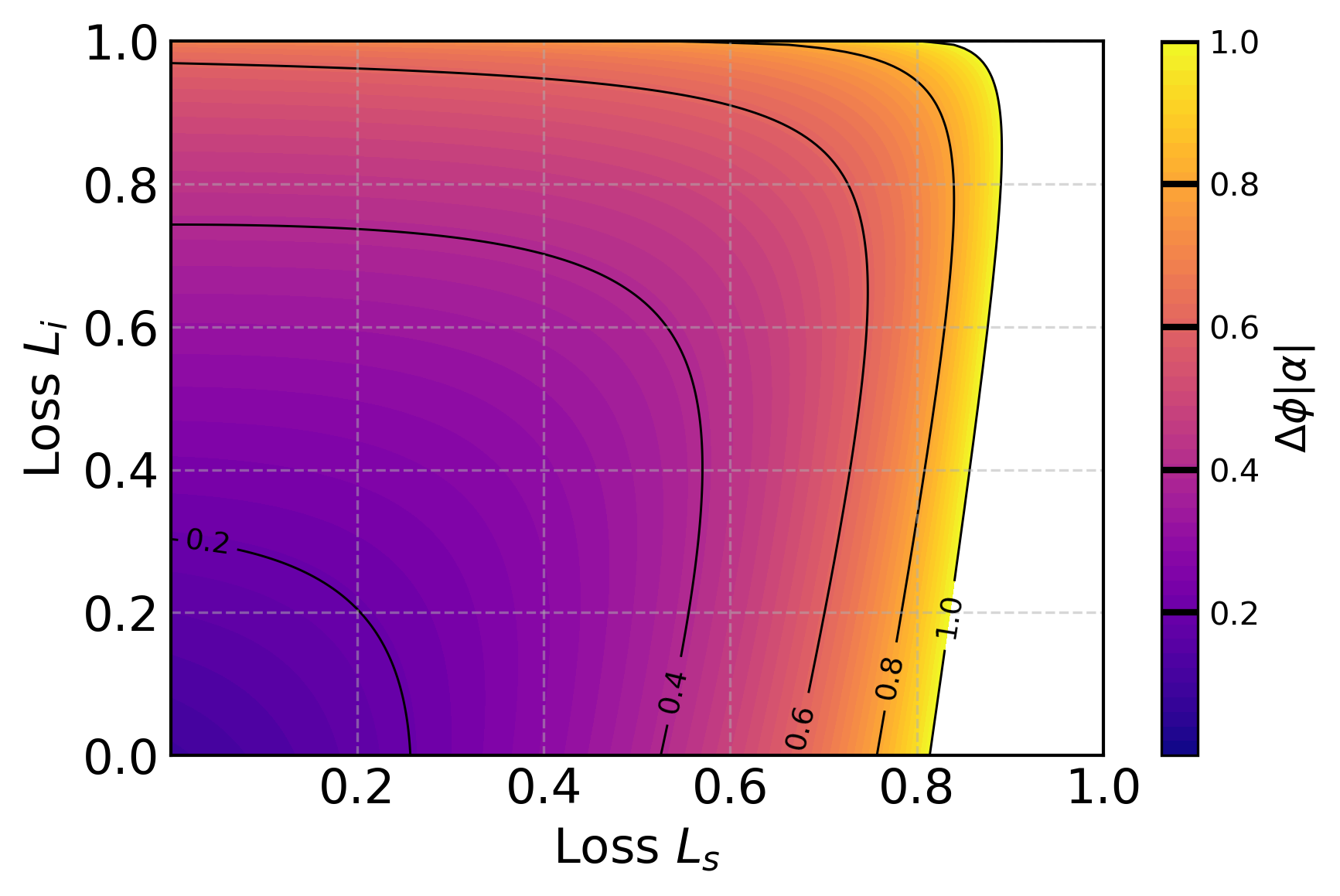} 
\vspace{2pt}
\textbf{(c)} 
\caption{\label{Fig04} Contour plot of the phase sensitivity versus losses in both arms of the SU(1,1) interferometer for (a) single-port detection with balanced gains ($G=2$), (b) single-port detection with unbalanced gains ($G_1=2,G_2=5$), (c) joint homodyne detection with balanced gains ($G=2$). }
\end{figure}
It is also interesting to observe the effect on the sensitivity of the variation of the losses in one arm while the losses in the other arm are kept constant. This is performed in Fig.\,\ref{Fig03}. Comparing Fig.\,\ref{Fig03}(a) and Fig.\,\ref{Fig03}(b) shows that for all three scenarios considered here, the losses degrade the sensitivity much more when they are located on the signal arm rather than on the idler arm. This degradation is so strong that for losses larger than 0.8, the SU(1,1) interferometer becomes less sensitive than a standard interferometer. Moreover, Fig.\,\ref{Fig03}(a) shows that while the scenario with balanced gains is the least sensitive one for $L_s=0$ and $L_i=0.2$, it becomes the most sensitive one when one increases   $L_s$ beyond a value of 0.3. The opposite behavior happens in Fig.\,\ref{Fig03}(b): while the balanced gain configuration is the most sensitive one for $L_s=0.2$ and $L_i=0$, it becomes the least sensitive one when one increases $L_i$ above $0.15$. These observations show that the optimization of the sensitivity of such an interferometer in the presence of losses is far from being intuitive.

 In a typical experiment, one should, of course, try to minimize the losses. However, for a given amount of losses on the two arms of the interferometer, it might be interesting to have a strategy to decide which of the three configurations considered here should be implemented to optimize the sensitivity. The aim of Fig.\,\ref{Fig04} is to try to determine such a strategy. This figure plots, in false colors, the evolution of the sensitivity normalized to $1/|\alpha|$, as a function of the losses $L_s$ and $L_i$. Figures \ref{Fig04}(a), \ref{Fig04}(b), and \ref{Fig04}(c) correspond respectively to the case of balanced gain with single-port homodyne detection, to the case of unbalanced gains with single-port homodyne detection, and to the case of balanced gains with joint homodyne detection. 
 
 One first notices that, remarkably, the sensitivity in the three configurations remains well below the sensitivity of an ideal MZI, even when losses in both arms are very high, and even though we consider relatively small gains. The second important observation is that the three schemes do not exhibit the same sensitivity to the internal losses, as already suggested by the results of Fig.\,\ref{Fig03}. This implies that, depending on the values of the losses $L_s$ and $L_i$, one needs to change the strategy in terms of gain balance and detection scheme. 

\begin{figure}[htp]
\centering
\includegraphics[width=0.45\textwidth]{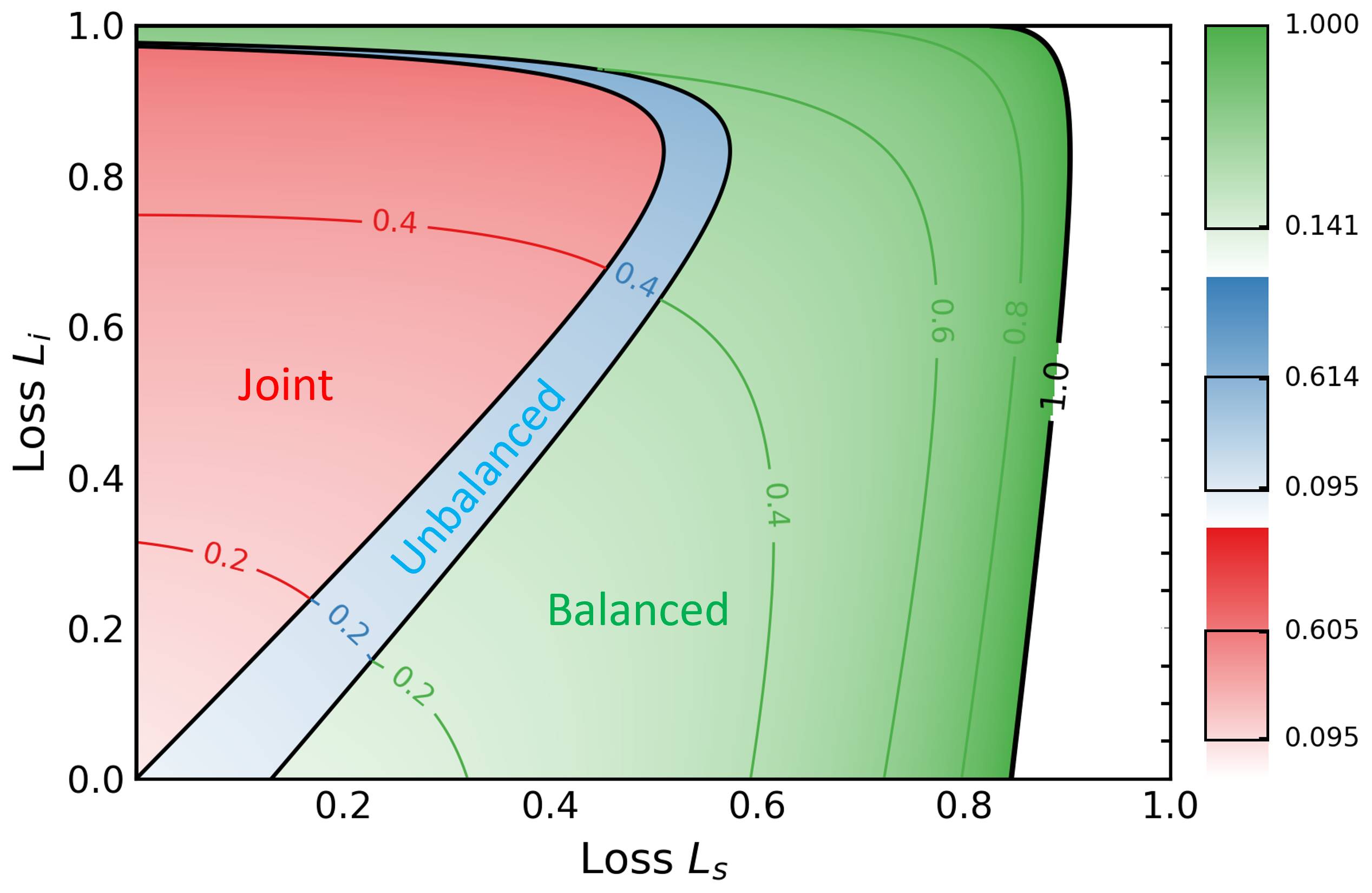} 
\caption{\label{Fig05} Choice of the gain and detection configurations leading to the best sensitivity for varying values of the losses $L_s$ and $L_i$. In each region, the displayed color corresponds to the best configuration. Green: single-port detection with balanced gains ($G=2$); Blue: single-port detection with unbalanced gains ($G_1=2, G_2=5$); Red: joint homodyne detection with balanced gains ($G=2$). The color becomes lighter when the sensitivity improves.
This plot is based on the data of Fig.\,\ref{Fig04}. All values are normalized to the case of the Mach-Zehnder interferometer probed with the same input state. Black boxes around color bars show the range of values used in the plot.} 
\end{figure}

 This conclusion is summarized in Fig.\,\ref{Fig05}. This figure combines the analytical results of the three plots of Fig.\,\ref{Fig04} and gives, for each couple of values of $L_s$ and $L_i$, the configuration that provides the best sensitivity. Remarkably, this figure shows that the joint homodyne detection, which is the most cumbersome one to implement experimentally, is a good choice only when the losses are relatively small. As soon as the losses become important, one should choose to implement single-port homodyne detection instead. Finally, only a small range of parameters leads to the choice of unbalanced gains, as shown by the narrow blue region in Fig.\,\ref{Fig05}.

 Similarly to the internal-loss-dependent sensitivity analysis, we also investigate the impact of external losses for a $\ket{\alpha,0}_{s,i}$ input state by applying the general framework developed in section\,\ref{sen exp}. Here, we highlight only the key observations, while the analytical expressions for the single-port homodyne at balanced and unbalanced gain configurations with arbitrary losses and for the joint homodyne with identical losses in a balanced-gain situation are provided in Table\,\ref{table3} of Appendix\,\ref{sen table}.  We find that single-port detection, under both balanced- and unbalanced-gain configurations, achieves better sensitivity than an MZI even in the presence of substantial external losses. Moreover, the unbalanced-gain configuration is more resilient to external losses than the balanced-gain case, since the gain of the second parametric amplifier can compensate for detection-side losses, in agreement with earlier studies. However, the influence of external losses on joint detection is markedly different at the bright- and dark-fringe operating points. While both $Y_{-}$ at the dark fringe and $Y_{+}$ at the bright fringe exhibit identical sensitivity enhancement relative to an MZI in the lossless balanced-gain limit, this advantage deteriorates rapidly at the bright fringe because external losses degrade the squeezing of the $Y_{+}$ quadrature, which serves as the primary resource for sensitivity enhancement. In contrast, the enhancement associated with the $Y_{-}$ quadrature at the dark fringe originates primarily from amplification of the phase-sensing signal rather than quadrature squeezing, making it substantially more robust against external losses. Consequently, dark-fringe joint detection offers superior resilience and represents a practically favorable operating regime under realistic external-loss conditions. Importantly, this detailed analysis, together with its clearer physical interpretation of external-loss effects in joint detection, constitutes a new result. All results related to external losses are summarized in Fig.\,\ref{fig: external_loss} of Appendix\,\ref{sen table}. 
 
\section{Conclusion}

We have developed a comprehensive theoretical framework that allows analytical evaluation of phase sensitivity for an SU(1,1) interferometric setup with arbitrary input states \textit{under homodyne detection}. A key result of our analysis is the formal equivalence between the full SU(1,1) interferometer (with two PAs) and its truncated counterpart, holding for any pair of input states. Our framework also enables analytical treatment of phase sensitivity in the presence of arbitrary internal losses. Notably, the sensitivity depends only on the quadrature statistics of the input states, making the approach general and independent of specific state representations. For instance, when operated in an unbalanced gain configuration, optimal sensitivity is achieved when the two input states have equal quadrature variances. In contrast, under balanced operation, the sensitivity becomes independent of the quadrature variance of one of the inputs. These insights provide valuable guidance for tailoring probe states to optimize performance.

As an application, we performed a detailed analysis of the case where one feeds the interferometer inputs with one coherent state in one arm and vacuum in the other arm.  Our results in this case show that single-port balanced homodyne detection exhibits strong robustness against internal losses, achieving sensitivity beyond the SQL even with substantial internal losses. Moreover, we have provided a comparison of the robustness of several configurations to internal losses, both balanced and unbalanced. This has allowed us to propose a helpful strategy to determine the best configuration for a given amount of losses. These results are prone to have an impact on the experimental implementations in which only one input laser is required.

In addition, we investigated the influence of external losses and demonstrated the high loss resilience of single-port homodyne detection while maintaining sensitivity beyond the SQL, with the unbalanced-gain configuration providing further robustness through compensation by the second parametric amplifier. For joint homodyne detection, the enhancement at the dark fringe remains significantly more robust than at the bright fringe, establishing dark fringe operation as the practically favorable regime under realistic external-loss conditions. 

Overall, this work confirms the interest of the SU(1,1) interferometer as a promising platform for surpassing the  SQL of linear interferometers in actual practical implementations. In future work, we aim to extend our framework using quantum Fisher information in order to determine whether homodyne detection achieves the ultimate sensitivity bound for arbitrary input states.

\begin{acknowledgments}
Sonu Jana acknowledges the financial support received from the Ministry of Education (MoE), Government of India, through the Prime Minister’s Research Fellowship (PMRF) program. Syamsundar De acknowledges the financial support received from the Ministry of Electronics and Information Technology (MeitY), Government of India, under the project “National Center for Quantum Accelerated Chips using Lithium Niobate (NCQAC).” This work was supported by the Physics Graduate School of Universit\'e Paris-Saclay (project NORI-MAQI) and the Institut des Sciences de la Lumière of Universit\'e Paris-Saclay (project ROLLMOPS). Fabien Bretenaker acknowledges support from IIT Kharagpur in the form of repeated invitations and an adjunct professor position.
* These authors contributed equally. 
\end{acknowledgments}

\bibliographystyle{unsrtnat}
\bibliography{apssamp}

\onecolumn\newpage
\appendix
In this appendix, we derive the sensitivity expressions given in the main body of the paper. Appendix \ref{transformations} provides the equation for the interferometer transformation in the case of ideal (i.e., lossless) and the case in the presence of loss. Appendix \ref{evolution} provides the output quadrature in terms of the input quadrature, both in the lossless case and in the presence of loss. Appendix \ref{statistic} provides the output quadrature statistic in both cases. Appendix \ref{sensitivity} presents the resulting sensitivity expression.  In Appendix \ref{Sensitivity enhancement}, we discuss the sensitivity enhancement mechanism under joint homodyne detection for both the full and truncated interferometers. Finally, Appendix \ref{sen table} presents all the sensitivity expressions derived in this article in a Formulary for ease of comparison.

\section{Interferometer transformations}
\label{transformations}
We obtain the interferometer transformations simply by applying the individual transformations associated with each element of the interferometer in the correct order. 
\subsection{Ideal SU(1,1) interferometer}
For the case of the ideal full SU(1,1) interferometer, this transformation is given by 
\begin{eqnarray}
    \begin{pmatrix}
    a_s^{(o)} \\
    a_i^{\dagger(o)}
\end{pmatrix} = \begin{pmatrix}
        G_2 & g_2 \\
        g_2 & G_2 \\
    \end{pmatrix}\begin{pmatrix}
        e^{i\phi} & 0 \\
        0 & 1 \\
    \end{pmatrix}\begin{pmatrix}
        G_1 & g_1 \\
        g_1 & G_1 \\
    \end{pmatrix}
     \begin{pmatrix}
    a_s^{(i)} \\
    a_i^{\dagger(i)}
\end{pmatrix}.
\end{eqnarray}
This effective transformation is then given by
\begin{eqnarray}\label{SU(1,1)_trans}
    \begin{pmatrix}
    a_s^{(o)} \\
    a_i^{\dagger(o)}
\end{pmatrix} = \begin{pmatrix}
        A & B \\
        D^* & C^* \\
    \end{pmatrix}
     \begin{pmatrix}
    a_s^{(i)} \\
    a_i^{\dagger(i)}
\end{pmatrix},
\end{eqnarray}
where $A = G_1G_2e^{i \phi} + g_1g_2$, $B = G_2g_1e^{i \phi} + G_1g_2$, $C = G_1G_2 + g_1g_2e^{-i \phi}$ and $D = G_2g_1 + G_1g_2e^{-i \phi}$. (see Eqs. \ref{SU(1,1)_transformation} in Section \ref{trnas coeff})  

\subsection{Ideal truncated SU(1,1) interferometer}
One may model the ideal truncated interferometer transformations in the same manner.
\begin{eqnarray}\label{Su 11 trnas truc}
    \begin{pmatrix}
    a_s^{(o)} \\
    a_i^{\dagger(o)}
\end{pmatrix} = \begin{pmatrix}
        e^{i\phi} & 0 \\
        0 & 1 \\
    \end{pmatrix}\begin{pmatrix}
        G & g \\
        g & G \\
    \end{pmatrix}
     \begin{pmatrix}
    a_s^{(i)} \\
    a_i^{\dagger(i)}
\end{pmatrix} \implies
\begin{pmatrix}
    a_s^{(o)} \\
    a_i^{\dagger(o)}
\end{pmatrix} = \begin{pmatrix}
        A & B \\
        D^* & C^* \\
    \end{pmatrix}
     \begin{pmatrix}
    a_s^{(i)} \\
    a_i^{\dagger(i)}
\end{pmatrix}.
\end{eqnarray}
Here, $A = Ge^{i\phi}$, $B = ge^{i\phi}$, $C = G$ and $D =g$.

\subsection{SU(1,1) interferometer with internal loss}
The transformation for the full interferometer with internal losses $L_i$ and $L_s$ (see fig \ref{Fig01} (a) in the main text) is given as
\begin{align}
\begin{pmatrix}
a_s^{(o)} \\
a_i^{\dagger(o)} \\
a_{sL}^{(o)} \\
a_{iL}^{\dagger(o)}
\end{pmatrix}
&=
\begin{pmatrix}
G_2 & g_2 & 0 & 0 \\
g_2 & G_2 & 0 & 0 \\
0 & 0 & 1 & 0 \\
0 & 0 & 0 & 1
\end{pmatrix}
\begin{pmatrix}
\sqrt{1-L_s} & 0 & \sqrt{L_s} & 0 \\
0 & \sqrt{1-L_i} & 0 & \sqrt{L_i} \\
\sqrt{L_s} & 0 & \sqrt{1-L_s} & 0 \\
0 & \sqrt{L_i} & 0 & \sqrt{1-L_i}
\end{pmatrix}\nonumber\\
&\times \begin{pmatrix}
e^{i\phi} & 0 & 0 & 0 \\
0 & 1 & 0 & 0 \\
0 & 0 & 1 & 0 \\
0 & 0 & 0 & 1
\end{pmatrix}
\begin{pmatrix}
G_1 & g_1 & 0 & 0 \\
g_1 & G_1 & 0 & 0 \\
0 & 0 & 1 & 0 \\
0 & 0 & 0 & 1
\end{pmatrix}
\begin{pmatrix}
a_s^{(i)} \\
a_i^{\dagger(i)} \\
a_{sL}^{(i)} \\
a_{iL}^{\dagger(i)}
\end{pmatrix},
\end{align}
using the extra modes $a_{sL}$ and $a_{iL}^{\dagger}$ as lossy modes that are added to model the noise/loss maybe not necessary. The effective transformation is then given as
\begin{eqnarray}
    \begin{pmatrix}
    a_s^{(o)} \\
    a_i^{\dagger(o)} \\
    a_{sL}^{(o)} \\
    a_{iL}^{\dagger(o)} \\
\end{pmatrix} = \begin{pmatrix}
        A_L & B_L & A'_L & B'_L \\
        D_L^* & C_L^* & D'^*_L & C'^*_L\\
        I'_L & J'_L & I_L & J_L\\
        M'^*_L & N'^*_L & M^*_L & N^*_L\\
    \end{pmatrix}
     \begin{pmatrix}
    a_s^{(i)} \\
    a_i^{\dagger(i)} \\
    a_{sL}^{(i)} \\
    a_{iL}^{\dagger(i)} \\
\end{pmatrix}. 
\end{eqnarray}
Here, $A_L = G_1G_2\sqrt{1-L_s}e^{i\phi} + g_1g_2\sqrt{1-L_i}$, $B_L = G_2g_1\sqrt{1-L_s}e^{i\phi} + G_1g_2\sqrt{1-L_i}$, $C_L = G_1G_2\sqrt{1-L_i} + g_1g_2\sqrt{1-L_s}e^{-i\phi}$, $D_L = G_2g_1\sqrt{1-L_i} + G_1g_2\sqrt{1-L_s}e^{-i\phi}$, $A_L' = G_2\sqrt{L_s}$, $B_L' = g_2\sqrt{L_i}$, $C_L' = G_2\sqrt{L_i}$ and $D_L' = g_2\sqrt{L_s}$. Since the lossy modes are irrelevant for sensitivity calculation, we do not provide the explicit expressions for the other constants.

\subsection{SU(1,1) interferometer with external losses}
The transformation for the full interferometer with external loss $l_s$ and $l_i$ (see fig \ref{Fig01}(a) is given as 
\begin{eqnarray}
    \begin{pmatrix}
    a_s^{(o)} \\
    a_i^{\dagger(o)} \\
    a_{sl}^{(o)} \\
    a_{il}^{\dagger(o)} \\
\end{pmatrix}& = &\begin{pmatrix}
        \sqrt{1-l_s} & 0 & \sqrt{l_s} & 0 \\
        0 & \sqrt{1-l_i} & 0 & \sqrt{l_i} \\
        \sqrt{l_s} & 0 & \sqrt{1-l_s} & 0\\
        0 & \sqrt{l_i} & 0 & \sqrt{1-l_i}\\
    \end{pmatrix}\begin{pmatrix}
        G_2 & g_2 & 0 & 0 \\
        g_2 & G_2 & 0 & 0\\
        0 & 0 & 1 & 0\\
        0 & 0 & 0 & 1\\
    \end{pmatrix}\begin{pmatrix}
        e^{i\phi} & 0 & 0 & 0 \\
        0 & 1 & 0 & 0 \\
        0 & 0 & 1 & 0\\
        0 & 0 & 0 & 1\\
    \end{pmatrix}\nonumber\\&\times&\begin{pmatrix}
        G_1 & g_1 & 0 & 0 \\
        g_1 & G_1 & 0 & 0 \\
        0 & 0 & 1 & 0\\
        0 & 0 & 0 & 1\\
    \end{pmatrix}
     \begin{pmatrix}
    a_s^{(i)} \\
    a_i^{\dagger(i)} \\
    a_{sl}^{(i)} \\
    a_{il}^{\dagger(i)} \\
\end{pmatrix} ,
\end{eqnarray}
leading to
\begin{eqnarray}
    \begin{pmatrix}
    a_s^{(o)} \\
    a_i^{\dagger(o)} \\
    a_{sl}^{(o)} \\
    a_{il}^{\dagger(o)} \\
\end{pmatrix} = \begin{pmatrix}
        A_l & B_l & A'_l & B'_l \\
        D_l^* & C_l^* & D'^*_l & C'^*_l\\
        I'_l & J'_l & I_l & J_l\\
        M'^*_l & N'^*_l & M^*_l & N^*_l\\
    \end{pmatrix}
     \begin{pmatrix}
    a_s^{(i)} \\
    a_i^{\dagger(i)} \\
    a_{sl}^{(i)} \\
    a_{il}^{\dagger(i)} \\
\end{pmatrix}.
\end{eqnarray}
Here, $A_l = \sqrt{1-l_s}A$, $B_l = \sqrt{1-l_s}B$, $C_l = \sqrt{1-l_i}C$, $D_l = \sqrt{1-l_i}D$, $A_l' = \sqrt{l_s}$, $C_l' = \sqrt{l_i}$ and $B_l' = D_l' = 0$. Other constants are not provided as they are not useful for sensitivity calculations.

\subsection{Truncated interferometer with losses}
For the truncated interferometer, we model the losses as arising after the phase shift. The transformations for the lossy truncated interferometer will then be given as
\begin{eqnarray}
    \begin{pmatrix}
    a_s^{(o)} \\
    a_i^{\dagger(o)} \\
    a_{sl}^{(o)} \\
    a_{il}^{\dagger(o)} \\
\end{pmatrix} = \begin{pmatrix}
        \sqrt{1-l_s} & 0 & \sqrt{l_s} & 0 \\
        0 & \sqrt{1-l_i} & 0 & \sqrt{l_i} \\
        \sqrt{l_s} & 0 & \sqrt{1-l_s} & 0\\
        0 & \sqrt{l_i} & 0 & \sqrt{1-l_i}\\
    \end{pmatrix}\begin{pmatrix}
        e^{i\phi} & 0 & 0 & 0 \\
        0 & 1 & 0 & 0 \\
        0 & 0 & 1 & 0\\
        0 & 0 & 0 & 1\\
    \end{pmatrix}\begin{pmatrix}
        G_1 & g_1 & 0 & 0 \\
        g_1 & G_1 & 0 & 0 \\
        0 & 0 & 1 & 0\\
        0 & 0 & 0 & 1\\
    \end{pmatrix}
     \begin{pmatrix}
    a_s^{(i)} \\
    a_i^{\dagger(i)} \\
    a_{sl}^{(i)} \\
    a_{il}^{\dagger(i)} \\
\end{pmatrix},
\end{eqnarray}

\begin{eqnarray}
    \begin{pmatrix}
    a_s^{(o)} \\
    a_i^{\dagger(o)} \\
    a_{sl}^{(o)} \\
    a_{il}^{\dagger(o)} \\
\end{pmatrix} = \begin{pmatrix}
        A_l & B_l & A'_l & B'_l \\
        D_l^* & C_l^* & D'^*_l & C'^*_l\\
        I'_l & J'_l & I_l & J_l\\
        M'^*_l & N'^*_l & M^*_l & N^*_l\\
    \end{pmatrix}
     \begin{pmatrix}
    a_s^{(i)} \\
    a_i^{\dagger(i)} \\
    a_{sl}^{(i)} \\
    a_{il}^{\dagger(i)} \\
\end{pmatrix}.
\end{eqnarray}
The relevant constants are given as $A_l = \sqrt{1-l_s}Ge^{i\phi}$, $B_l = \sqrt{1-l_s}ge^{i\phi}$, $C_l = \sqrt{1-l_i}G$, $D_l = \sqrt{1-l_i}g$, $A_l' = \sqrt{l_s}$, $C_l' = \sqrt{l_i}$ and $B_l' = D_l' = 0$. 

\section{Evolution of field quadratures}
\label{evolution}
In this section, we derive the expressions for the output port field quadrature observables in terms of the input field quadrature operator. 

\subsection{Ideal interferometer}
Note that since the form of the transformations for both the truncated and the ideal interferometer Eqs. (\ref{SU(1,1)_trans}) and (\ref{Su 11 trnas truc}) are analogous, the expressions we derive apply to both cases. One must then use the appropriate expressions for the A, B, C, D constants in each case. First, we state the expression for the field quadrature $Y$ at the first output port 
\begin{eqnarray}
    Y_s^{(o)} = \frac{a_s^{(o)} - a_s^{\dagger(o)}}{i} = \frac{(Aa_s^{(i)} + Ba_i^{\dagger(i)}) - (A^*a_s^{\dagger(i)} + B^*a_i^{(i)})}{i} \qquad \qquad \quad \nonumber \\ = |A|\frac{e^{i\kappa_A}a_s^{(i)} - e^{-i\kappa_A}a_s^{\dagger(i)}}{i} - |B|\frac{e^{-i\kappa_B}a_i^{(i)} - e^{i\kappa_B}a_i^{\dagger(i)}}{i} = |A|Y_s^{(i)}(\kappa_A) - |B|Y_i^{(i)}(-\kappa_B).
\end{eqnarray}

Similarly, we derive the expressions for the field quadratures at the second output port as follows
\begin{eqnarray}
    Y_i^{(o)} = \frac{a_i^{(o)} - a_i^{\dagger(o)}}{i} = \frac{(Ca_i^{(i)} + Da_s^{\dagger(i)}) - (C^*a_i^{\dagger(i)} + D^*a_i^{(i)})}{i} \qquad \qquad \quad \nonumber \\ = |C|\frac{e^{i\kappa_C}a_i^{(i)} - e^{-i\kappa_C}a_i^{\dagger(i)}}{i} - |D|\frac{e^{-i\kappa_D}a_s^{(i)} - e^{i\kappa_D}a_s^{\dagger(i)}}{i} = |C|Y_i^{(i)}(\kappa_C) - |D|Y_s^{(i)}(-\kappa_D)
\end{eqnarray}

Lastly, we establish the evolution of the joint (sum and difference) homodyne measurements.
\begin{eqnarray}
    Y_{\pm}^{(o)} = \frac{a_s^{(o)} - a_s^{\dagger(o)}}{i} \pm \frac{a_i^{(o)} - a_i^{\dagger(o)}}{i} = \frac{(Aa_s^{(i)} + Ba_i^{\dagger(i)}) - (A^*a_s^{\dagger(i)} + B^*a_i^{(i)}) \pm (Ca_i^{(i)} + Da_s^{\dagger(i)}) \mp (C^*a_i^{\dagger(i)} + D^*a_s^{(i)})}{i} \nonumber \\ = |A \mp D^*| \frac{e^{i\kappa_{A \mp D^*}}a_s^{(i)} - e^{-i\kappa_{A \mp D^*}}a_s^{\dagger(i)}}{i} + |C \mp B^*|\frac{e^{i\kappa_{C \mp B^*}}a_i^{(i)} - e^{-i\kappa_{C \mp B^*}}a_i^{\dagger(i)}}{i} \qquad \qquad \qquad \qquad \nonumber \\ = |A \mp D^*|Y_s^{(i)}(\kappa_{A \mp D^*}) + |C \mp B^*|Y_i^{(i)}(\kappa_{C \mp B^*}) = |E|Y_s^{(i)}(\kappa_E) + |F|Y_i^{(i)},(\kappa_F) \qquad \qquad \qquad \qquad
\end{eqnarray}

Here, the final equality serves as a definition of constants $E$ and $F$ used in the main manuscript. From here, the expressions for the sensitivities can be obtained in a straightforward manner, as we will demonstrate in the next section.

\subsection{Lossy interferometer}
As is the case for the ideal interferometer, all the expressions for the different lossy interferometers are formally identical and only differ in the constants. Here we use the notation $X_L$ for constants that have been used for internal noise in this study, but the results generalize to the case of a general (internal or external) lossy interferometer. \\
Now, let us evaluate the expression for the output port field quadrature $Y_s^{(o)}$.
\begin{eqnarray}
    Y_s^{(o)} &= &\frac{a_s^{(o)} - a_s^{\dagger(o)}}{i}\nonumber\\ &=& \frac{(A_La_s^{(i)} + B_La_i^{\dagger(i)} + A_L'a_{sL}^{(i)} + B_L'a_{iL}^{\dagger(i)}) - (A_L^*a_s^{\dagger(i)} + B_L^*a_i^{(i)} + A_L'^*a_{sL}^{\dagger(i)} + B_L'^*a_{iL}^{(i)})}{i}   \nonumber \\ &=& |A_L|\frac{e^{i\kappa_{A_L}}a_s^{(i)} - e^{-i\kappa_{A_L}}a_s^{\dagger(i)}}{i} - |B_L|\frac{e^{-i\kappa_{B_L}}a_i^{(i)} - e^{i\kappa_{B_L}}a_i^{\dagger(i)}}{i} + |A_L'|\frac{e^{i\kappa_{A_L'}}a_{sL}^{(i)} - e^{-i\kappa_{A_L'}}a_{sL}^{\dagger(i)}}{i} \nonumber\\&&- |B_L'|\frac{e^{-i\kappa_{B_L'}}a_{iL}^{(i)} - e^{i\kappa_{B_L'}}a_{iL}^{\dagger(i)}}{i} \nonumber \\ &= &|A_L|Y_s^{(i)}(\kappa_{A_L}) - |B_L|Y_i^{(i)}(-\kappa_{B_L}) + |A_L'|Y_{sL}^{(i)}(\kappa_{A_L'}) - |B_L'|Y_{iL}^{(i)}(-\kappa_{B_L'}). 
\end{eqnarray}

Similarly, we obtain the expression for the second port and joint port measurements as
\begin{eqnarray}
    Y_i^{(o)} = |C_L|Y_i^{(i)}(\kappa_{C_L}) - |D_L|Y_s^{(i)}(-\kappa_{D_L}) + |C_L'|Y_{iL}^{(i)}(\kappa_{C_L'}) - |D_L'|Y_{sL}^{(i)}(-\kappa_{D_L'}) 
\end{eqnarray}
\begin{eqnarray}
\label{YO}
    Y_{\pm}^{(o)} &= &|A_L \mp D_L^*|Y_s^{(i)}(\kappa_{A_L \mp D_L^*}) + |C_L \mp B_L^*|Y_i^{(i)}(\kappa_{C_L \mp B_L^*}) + |A_L' \mp D_L'^*|Y_{sL}^{(i)}(\kappa_{A_L' \mp D_L'^*}) \nonumber\\&&+ |C_L' \mp B_L'^*|Y_{iL}^{(i)}(\kappa_{C_L' \mp B_L'^*})  \nonumber \\ &= &|E_L|Y_s^{(i)}(\kappa_{E_L}) + |F_L|Y_i^{(i)}(\kappa_{F_L}) + |E_L'|Y_{sL}^{(i)}(\kappa_{E_L'}) + |F_L'|Y_{iL}^{(i)}(\kappa_{F_L'}) .
\end{eqnarray}
The final equality defines the constants $E_L, F_L, E_L'$, and $F_L'$ used in the main manuscript. Note that this equation (\ref{YO})  applies for the case of internal and external noise with an appropriate choice of index ($L$ or $l$).

\section{Output port statistics}
\label{statistic}
In this section, we derive the expressions for the output port statistics as a function of the input state statistics. The results of this section follow straightforwardly from the previous section. 
\subsection{Ideal interferometer}
For the single-port measurement at the first output port we have the following expression.
\begin{eqnarray}
    \braket{Y_s^{(o)}} = |A|\braket{Y_s^{(i)}(\kappa_A)} - |B|\braket{Y_i^{(i)}(-\kappa_B)},
\end{eqnarray}
\begin{eqnarray}
    \braket{(Y^2_s)^{(o)}} = |A|^2\braket{(Y_s^2(\kappa_A))^{(i)}} + |B|^2\braket{(Y_i^2(-\kappa_B))^{(i)}} - 2|A||B|\braket{Y_s^{(i)}(\kappa_A)Y_i^{(i)}(-\kappa_B)},
\end{eqnarray}
and
\begin{eqnarray}\label{outvar}
    (\Delta Y_s^{(o)})^2 &=& |A|^2(\Delta Y_s^{(i)}(\kappa_A))^2 + |B|^2(\Delta Y_i^{(i)}(-\kappa_B))^2 \nonumber\\&&- 2|A||B|(\braket{Y_s^{(i)}(\kappa_A)Y_i^{(i)}(-\kappa_B)} - \braket{Y_s^{(i)}(\kappa_A)}\braket{Y_i^{(i)}(-\kappa_B)}). \qquad
\end{eqnarray}

Similarly, we evaluate the statistics of the second output port as 
\begin{eqnarray}
    \braket{Y_i^{(o)}} = |C|\braket{Y_i^{(i)}(\kappa_C)} - |D|\braket{Y_s^{(i)}(-\kappa_D)},
\end{eqnarray}
\begin{eqnarray}
    \braket{(Y^2_i)^{(o)}} = |C|^2\braket{(Y_i^2(\kappa_C))^{(i)}} + |D|^2\braket{(Y_s^2(-\kappa_D))^{(i)}} - 2|C||D|\braket{Y_i^{(i)}(\kappa_C)Y_s^{(i)}(-\kappa_D)},
\end{eqnarray}
and
\begin{eqnarray}
    (\Delta Y_i^{(o)})^2 &=& |C|^2(\Delta Y_i^{(i)}(\kappa_C))^2 + |D|^2(\Delta Y_s^{(i)}(-\kappa_D))^2 \nonumber\\
    &&- 2|C||D|(\braket{Y_i^{(i)}(\kappa_C)Y_s^{(i)}(-\phi_D)} - \braket{Y_i^{(i)}(\kappa_C)}\braket{Y_s^{(i)}(-\kappa_D)}). \qquad
\end{eqnarray}
Finally, we state the joint measurement statistics.
\begin{eqnarray}
    \braket{Y_{\pm}^{(o)}} = |E|\braket{Y_s^{(i)}(\kappa_E)} + |F|\braket{Y_i^{(i)}(\kappa_F)},
\end{eqnarray}
\begin{eqnarray}
    \braket{(Y^2_{\pm})^{(o)}} = |E|^2\braket{(Y_s^2(\kappa_E))^{(i)}} + |F|^2\braket{(Y_i^2(\kappa_F))^{(i)}} + 2|E||F|\braket{Y_s^{(i)}(\kappa_E)Y_i^{(i)}(\kappa_F)},
\end{eqnarray}
and
\begin{eqnarray}\label{Joint variance}
    (\Delta Y_{\pm}^{(o)})^2 &=& |E|^2(\Delta Y_s^{(i)}(\kappa_E))^2 + |F|^2(\Delta Y_i^{(i)}(\kappa_F))^2 \nonumber\\
    &&+ 2|E||F|(\braket{Y_s^{(i)}(\kappa_E)Y_i^{(i)}(\kappa_F)} - \braket{Y_s^{(i)}(\kappa_E)}\braket{Y_i^{(i)}(\kappa_F)}).
\end{eqnarray}

\subsection{Lossy interferometer}
The expressions for the noisy interferometers are obtained in the same manner as in the previous section. \\
The expressions for the single-port statistics of a noisy SU(1,1) interferometer for the first and second output port respectively are given as follows.
\begin{eqnarray}
    \braket{Y_s^{(o)}} &=& |A_L|\braket{Y_s^{(i)}(\kappa_{A_L})} - |B_L|\braket{Y_i^{(i)}(-\kappa_{B_L})} + |A_L'|\braket{Y_{sL}^{(i)}(\kappa_{A_L'})} - |B_L'|\braket{Y_{iL}^{(i)}(-\kappa_{B_L'})} \nonumber \\ &=& |A_L|\braket{Y_s^{(i)}(\kappa_{A_L})} - |B_L|\braket{Y_i^{(i)}(-\kappa_{B_L})}
\end{eqnarray} 
Note, that we obtain the final expression because the inputs of the lossy modes are taken to be vacuum. This produces the expected result of reducing the signal arising from the input probe with no contribution from the lossy modes. As we will show, the lossy modes do, however, contribute to the output noise.
\begin{align}
\braket{(Y_s^2)^{(o)}} 
&= |A_L|^2 \braket{ \left( Y_s^2(\kappa_{A_L}) \right)^{(i)} }
 + |B_L|^2 \braket{ \left( Y_i^2(-\kappa_{B_L}) \right)^{(i)} } \nonumber \\
&\quad
 - 2 |A_L| |B_L|
 \braket{ Y_s^{(i)}(\kappa_{A_L}) \, Y_i^{(i)}(-\kappa_{B_L}) } \nonumber \\
&\quad
 + |A_L'|^2 \braket{ \left( Y_s^2(\kappa_{A_L'}) \right)^{(i)} }
 + |B_L'|^2 \braket{ \left( Y_i^2(-\kappa_{B_L'}) \right)^{(i)} } \nonumber \\
&\quad
 - 2 |A_L'| |B_L'|
 \braket{ Y_s^{(i)}(\kappa_{A_L'}) \, Y_i^{(i)}(-\kappa_{B_L'}) } .
\end{align}

Again, here we use the fact that the input of the lossy modes is the vacuum.
\begin{eqnarray}
    (\Delta Y_s^{(o)})^2 &=& |A_L|^2(\Delta Y_s^{(i)}(\kappa_{A_L}))^2 + |B_L|^2(\Delta Y_i^{(i)}(-\kappa_{B_L}))^2  \nonumber \\ &&- 2|A_L||B_L|(\braket{Y_s^{(i)}(\kappa_{A_L})Y_i^{(i)}(-\kappa_{B_L})} - \braket{Y_s^{(i)}(\kappa_{A_L})}\braket{Y_i^{(i)}(-\kappa_{B_L})}) \nonumber \\ && + |A_L'| + |B_L'|
\end{eqnarray}

Similarly, the expressions for the statistics at the second output port are given by
\begin{eqnarray}
    \braket{Y_i^{(o)}} &=& |C_L|\braket{Y_i^{(i)}(\kappa_{C_L})} - |D_L|\braket{Y_s^{(i)}(-\kappa_{D_L})} + |C_L'|\braket{Y_{iL}^{(i)}(\kappa_{C_L'})} - |D_L'|\braket{Y_{sL}^{(i)}(-\kappa_{D_L'})} \nonumber \\ &=& |C_L|\braket{Y_i^{(i)}(\kappa_{C_L})} - |D_L|\braket{Y_s^{(i)}(-\kappa_{D_L})}
\end{eqnarray}
\begin{eqnarray}
    \braket{(Y^2_i)^{(o)}} &= &|C_L|^2\braket{(Y_i^2(\kappa_{C_L}))^{(i)}} + |D_L|^2\braket{(Y_s^2(-\kappa_{D_L}))^{(i)}} - 2|C_L||D_L|\braket{Y_i^{(i)}(\kappa_{C_L})Y_i^{(i)}(-\kappa_{D_L})} \nonumber \\ 
    &&+ |C_L'|^2\braket{(Y_i^2(\kappa_{C_L'}))^{(i)}} + |D_L'|^2\braket{(Y_s^2(-\kappa_{D_L'}))^{(i)}} - 2|C_L'||D_L'|\braket{Y_i^{(i)}(\kappa_{C_L'})Y_s^{(i)}(-\kappa_{D_L'})} \nonumber \\ 
    &=& |C_L|^2\braket{(Y_i^2(\kappa_{C_L}))^{(i)}} + |D_L|^2\braket{(Y_s^2(-\kappa_{D_L}))^{(i)}}\nonumber\\
    &&- 2|C_L||D_L|\braket{Y_i^{(i)}(\kappa_{C_L})Y_s^{(i)}(-\kappa_{D_L})} + |C_L'| + |D_L'|
\end{eqnarray}
\begin{eqnarray}
    (\Delta Y_i^{(o)})^2 &=& |C_L|^2(\Delta Y_i^{(i)}(\kappa_{C_L}))^2 + |D_L|^2(\Delta Y_i^{(i)}(-\kappa_{D_L}))^2 \nonumber \\ 
    &&- 2|C_L||D_L|(\braket{Y_i^{(i)}(\kappa_{C_L})Y_s^{(i)}(-\kappa_{D_L})} - \braket{Y_i^{(i)}(\kappa_{C_L})}\braket{Y_s^{(i)}(-\kappa_{D_L})}) + |C_L'| + |D_L'|
\end{eqnarray}

Finally, we state the statistics of the joint homodyne measurements.
\begin{eqnarray}
    \braket{Y_{\pm}^{(o)}} &=& |E_L|\braket{Y_s^{(i)}(\kappa_{A_L})} + |F_L|\braket{Y_i^{(i)}(\kappa_{F_L})} + |E_L'|\braket{Y_{sL}^{(i)}(\kappa_{E_L'})} + |F_L'|\braket{Y_{iL}^{(i)}(\kappa_{F_L'})} \nonumber \\ 
    &=& |E_L|\braket{Y_s^{(i)}(\kappa_{E_L})} + |F_L|\braket{Y_i^{(i)}(\kappa_{F_L})},
\end{eqnarray}
\begin{eqnarray}
    \braket{(Y^2_{\pm})^{(o)}} &=& |E_L|^2\braket{(Y_s^2(\kappa_{E_L}))^{(i)}} + |F_L|^2\braket{(Y_i^2(\kappa_{F_L}))^{(i)}} + 2|E_L||F_L|\braket{Y_s^{(i)}(\kappa_{E_L})Y_i^{(i)}(\kappa_{F_L})} \nonumber \\ 
    &&+ |E_L'|^2\braket{(Y_s^2(\kappa_{E_L'}))^{(i)}} + |F_L'|^2\braket{(Y_i^2(\kappa_{F_L'}))^{(i)}} + 2|E_L'||F_L'|\braket{Y_s^{(i)}(\kappa_{E_L'})Y_i^{(i)}(\kappa_{F_L'})} \nonumber \\ 
    &=& |E_L|^2\braket{(Y_s^2(\kappa_{E_L}))^{(i)}} + |F_L|^2\braket{(Y_i^2(\kappa_{F_L}))^{(i)}} \nonumber\\
    &&+ 2|E_L||F_L|\braket{Y_s^{(i)}(\kappa_{E_L})Y_i^{(i)}(\kappa_{F_L})} + |E_L'| + |F_L'|,
\end{eqnarray}
\begin{eqnarray}
    (\Delta Y_{\pm}^{(o)})^2 &=& |E_L|^2(\Delta Y_s^{(i)}(\kappa_{E_L}))^2 + |F_L|^2(\Delta Y_i^{(i)}(\kappa_{F_L}))^2 \nonumber \\ 
    &&+ 2|E_L||F_L|(\braket{Y_s^{(i)}(\kappa_{E_L})Y_i^{(i)}(\kappa_{F_L})} - \braket{Y_s^{(i)}(\kappa_{E_L})}\braket{Y_i^{(i)}(\kappa_{F_L})}) + |E_L'| + |F_L'|
\end{eqnarray}

\section{Sensitivity expressions}
\label{sensitivity}

\subsection{Ideal interferometers}

\subsubsection{Full SU(1,1) interferometer}
Now we evaluate the signal at the first output port, associated with the phase shift.
\begin{eqnarray}\label{phasechangesignal}
    \frac{\partial \braket{Y_s^{(o)}}}{\partial \phi} &=& \frac{\partial\braket{(Aa_s^{(i)} - A^*a_s^{\dagger(i)})}}{i\partial \phi} - \frac{\partial\braket{(B^*a_i^{(i)} - Ba_i^{\dagger(i)})}}{i\partial \phi}     \nonumber \\ 
    &=& G_2G_1\braket{e^{i\phi}a_s^{(i)} + e^{-i\phi}a_s^{\dagger(i)}} + G_2g_1\braket{e^{-i\phi}a_i^{(i)} + e^{i\phi}a_i^{\dagger(i)}}\nonumber\\
    &=& G_2(G_1\braket{X_s^{(i)}(\phi)} + g_1\braket{X_i^{(i)}(-\phi)})
\end{eqnarray}
Finally, we have the expression for sensitivity as
\begin{eqnarray}
    (\Delta \phi)_{Y_s} = \frac{\sqrt{|A|^2(\Delta Y_s^{(i)}(\kappa_A))^2 + |B|^2(\Delta Y_i^{(i)}(-\kappa_B))^2 - 2|A||B|(\braket{Y_s^{(i)}(\kappa_A)Y_i^{(i)}(-\kappa_B)} - \braket{Y_s^{(i)}(\kappa_A)}\braket{Y_i^{(i)}(-\kappa_B)})}}{G_2|G_1\braket{X_s^{(i)}(\phi)} + g_1\braket{X_i^{(i)}(-\phi)}|}.\nonumber\\
\end{eqnarray}
At the dark fringe, in both balanced and unbalanced operation, $\kappa_A = (2n+1)\pi$ and $\phi_B = 2n\pi$. This implies that $Y(\kappa_A) = -Y$ and $Y(\phi_B) = Y$. Also, note that $X(\phi) = X(-\phi) = -X$. The resulting sensitivity expression then reduces to
\begin{eqnarray}
    (\Delta \phi)_{Y_s} = \frac{\sqrt{|A|^2(\Delta Y_s^{(i)})^2 + |B|^2(\Delta Y_i^{(i)})^2 + 2|A||B|(\braket{Y_s^{(i)}Y_i^{(i)}} - \braket{Y_s^{(i)}}\braket{Y_i^{(i)}})}}{G_2|G_1\braket{X_s^{(i)}} + g_1\braket{X_i^{(i)}}|}.
\end{eqnarray}

The expression for the signal associated to the phase change at the second output port is 
\begin{eqnarray}
  \frac{\partial \braket{Y_i^{(o)}}}{\partial \phi} &= &\frac{\partial\braket{(Ca_i^{(i)} - C^*a_i^{\dagger(i)})}}{i\partial \phi} - \frac{\partial\braket{(D^*a_s^{(i)} - Da_i^{\dagger(i)})}}{i\partial \phi}  \nonumber \\ 
  &=& -g_2g_1\braket{e^{-i\phi}a_i^{(i)} + e^{i\phi}a_i^{\dagger(i)}} - g_2G_1\braket{e^{i\phi}a_s^{(i)} + e^{-i\phi}a_s^{\dagger(i)}}\nonumber\\
  &=& -g_2(G_1\braket{X_s^{(i)}(\phi)} + g_1\braket{X_i^{(i)}(-\phi)}).  
\end{eqnarray}

Also note that at the dark fringe, $\kappa_C = 2n\pi$  and $\kappa_D = (2n+1)\pi$ for both operating conditions. The resulting sensitivity expression is given as
\begin{eqnarray}
    (\Delta \phi)_{Y_i} = \frac{\sqrt{|C|^2(\Delta Y_i^{(i)})^2 + |D|^2(\Delta Y_s^{(i)})^2 + 2|C||D|(\braket{Y_i^{(i)}Y_s^{(i)}} - \braket{Y_i^{(i)}}\braket{Y_s^{(i)}})}}{g_2|G_1\braket{X_s^{(i)}} + g_1\braket{X_i^{(i)}}|}.
\end{eqnarray}

Finally, we state the expression for the signal associated with the phase change for the joint homodyne measurements. The expression directly follows from the expressions above as
\begin{eqnarray}\label{phase change signal full}
     \frac{\partial \braket{Y_{\pm}^{(o)}}}{\partial \phi} = \frac{\partial \braket{Y_s^{(o)} \pm Y_2^{(o)}}} {\partial \phi} = (G_2 \mp g_2)(G_1\braket{X_s^{(i)}(\phi)} + g_1\braket{X_i^{(i)}(-\phi)}).
\end{eqnarray}

The resulting expression for sensitivity, at both the bright and dark fringes, is given as
\begin{eqnarray}
    (\Delta \phi)_{Y_{\pm}} = \frac{\sqrt{|E|^2(\Delta Y_s^{(i)})^2 + |F|^2(\Delta Y_i^{(i)})^2 + 2|E||F|(\braket{Y_s^{(i)}Y_i^{(i)}} - \braket{Y_s^{(i)}}\braket{Y_i^{(i)}})}}{(G_2 \mp g_2)|G_1\braket{X_s^{(i)}} + g_1\braket{X_i^{(i)}}|}.
\end{eqnarray}

\subsubsection{Truncated interferometer}
In the truncated interferometer, since only the top arm of the interferometer is modulated by the unknown phase, without a second PA to close the interference, the signal at the bottom arm is independent of the phase. Thus,
\begin{eqnarray}\label{phase change signal trun}
    \frac{\partial \braket{Y_{\pm}^{(o)}}}{\partial \phi} = \frac{\partial \braket{Y_s^{(o)}}}{\partial \phi} = \frac{\partial\braket{(Aa_s^{(i)} - A^*a_s^{\dagger(i)})}}{i\partial \phi} - \frac{\partial\braket{(B^*a_i^{(i)} - Ba_i^{\dagger(i)})}}{i\partial \phi} \qquad \quad \nonumber \\ = G\braket{e^{i\phi}a_s^{(i)} + e^{-i\phi}a_s^{\dagger(i)}} + g\braket{e^{-i\phi}a_i^{(i)} + e^{i\phi}a_i^{\dagger(i)}} = G\braket{X_s^{(i)}(\phi)} + g\braket{X_i^{(i)}(-\phi)}.
\end{eqnarray}
The expression for sensitivity at both the bright and dark fringes is then given as
\begin{eqnarray}
    (\Delta \phi)_{Y_{\pm}} = \frac{\sqrt{|E|^2(\Delta Y_s^{(i)})^2 + |F|^2(\Delta Y_i^{(i)})^2 + 2|E||F|(\braket{Y_s^{(i)}Y_i^{(i)}} - \braket{Y_s^{(i)}}\braket{Y_i^{(i)}})}}{|G\braket{X_s^{(i)}} + g\braket{X_i^{(i)}}|}.
\end{eqnarray}

\subsection{Lossy interferometers}
\subsubsection{Full SU(1,1) interferometer}
For the case of internal loss, the signal associated with the phase change for measurement at the first output port is given by
\begin{eqnarray}
    \frac{\partial \braket{Y_s^{(o)}}}{\partial \phi} &=& \frac{\partial\braket{(A_La_s^{(i)} - A_L^*a_s^{\dagger(i)})}}{i\partial \phi} - \frac{\partial\braket{(B_L^*a_i^{(i)} - B_La_i^{\dagger(i)})}}{i\partial \phi} \nonumber \\ 
    &= &\sqrt{1-L_s}G_2(G_1\braket{X_s^{(i)}(\phi)} + g_1\braket{X_i^{(i)}(-\phi)}) \qquad \quad
\end{eqnarray}
Note that this expression is a result of the fact that $\frac{\partial A_L'}{\partial \phi} = \frac{\partial B_L'}{\partial \phi} = 0$. The resulting expression for the sensitivity at the dark fringe is given as 
\begin{eqnarray}
\label{sens_Ys_losses}
    (\Delta \phi)_{Y_s} = \frac{\sqrt{|A_L|^2(\Delta Y_s^{(i)})^2 + |B_L|^2(\Delta Y_i^{(i)})^2 + 2|A_L||B_L|(\braket{Y_s^{(i)}Y_i^{(i)}} - \braket{Y_s^{(i)}}\braket{Y_i^{(i)}}) + |A_L'| + |B_L'|}}{\sqrt{1-L_s}G_2|G_1\braket{X_s^{(i)}} + g_1\braket{X_i^{(i)}}|}.\nonumber\\
\end{eqnarray}
Now we state the expressions for measurement at the second output port. 
\begin{eqnarray}
    \frac{\partial \braket{Y_i^{(o)}}}{\partial \phi} = \frac{\partial\braket{(C_La_i^{(i)} - C_L^*a_i^{\dagger(i)})}}{i\partial \phi} - \frac{\partial\braket{(D_L^*a_s^{(i)} - D_La_s^{\dagger(i)})}}{i\partial \phi} \nonumber \\ = -\sqrt{1-L_s}g_2(G_1\braket{X_s^{(i)}(\phi)} + g_1\braket{X_i^{(i)}(-\phi)}) \qquad \quad
\end{eqnarray}
At the dark fringe, the expression for sensitivity becomes
\begin{eqnarray}
    (\Delta \phi)_{Y_i} = \frac{\sqrt{|C_L|^2(\Delta Y_i^{(i)})^2 + |D_L|^2(\Delta Y_s^{(i)})^2 + 2|C_L||D_L|(\braket{Y_i^{(i)}Y_s^{(i)}} - \braket{Y_i^{(i)}}\braket{Y_s^{(i)}}) + |C_L'| + |D_L'|}}{\sqrt{1-L_s}g_2|G_1\braket{X_s^{(i)}} + g_1\braket{X_i^{(i)}}|}.\nonumber\\
\end{eqnarray}
From the expressions for single-port signals as given above, it is clear that
\begin{eqnarray}
    \frac{\partial \braket{Y_{\pm}^{(o)}}}{\partial \phi} = \sqrt{1-L_s}(G_2 \mp g_2)(G_1\braket{X_s^{(i)}(\phi)} + g_1\braket{X_i^{(i)}(-\phi)}), \qquad \quad
\end{eqnarray}
and sensitivity, at both the bright and dark fringes, is given as
\begin{eqnarray}
\label{sens_Yjoint_losses}
    (\Delta \phi)_{Y_{\pm}} = \frac{\sqrt{|E_L|^2(\Delta Y_s^{(i)})^2 + |F_L|^2(\Delta Y_i^{(i)})^2 + 2|E_L||F_L|(\braket{Y_s^{(i)}Y_i^{(i)}} - \braket{Y_s^{(i)}}\braket{Y_i^{(i)}}) + |E_L'| + |F_L'|}}{\sqrt{1-L_s}(G_2 \mp g_2)|G_1\braket{X_s^{(i)}} + g_1\braket{X_i^{(i)}}|}.\nonumber\\
\end{eqnarray}

Now we consider the expressions for full SU(1,1) interferometers with external loss. Firstly, the expressions for the phase change signal for measurement at the first output port is given as
\begin{eqnarray}
    \frac{\partial \braket{Y_s^{(o)}}}{\partial \phi} = \frac{\partial\braket{(A_la_s^{(i)} - A_l^*a_s^{\dagger(i)})}}{i\partial \phi} - \frac{\partial\braket{(B_l^*a_i^{(i)} - B_la_i^{\dagger(i)})}}{i\partial \phi} \nonumber \\ = \sqrt{1-l_s}G_2(G_1\braket{X_s^{(i)}(\phi)} + g_1\braket{X_i^{(i)}(-\phi)}) \qquad \quad
\end{eqnarray}
The resultant equation for sensitivity is then given as
\begin{eqnarray}
    (\Delta \phi)_{Y_s} = \frac{\sqrt{|A_l|^2(\Delta Y_s^{(i)})^2 + |B_l|^2(\Delta Y_i^{(i)})^2 + 2|A_l||B_l|(\braket{Y_s^{(i)}Y_i^{(i)}} - \braket{Y_s^{(i)}}\braket{Y_i^{(i)}}) + |A_l'| + |B_l'|}}{\sqrt{1-l_s}G_2|G_1\braket{X_s^{(i)}} + g_1\braket{X_i^{(i)}}|}.
\end{eqnarray}
The analogous expressions for the measurement at the second output port are then
\begin{eqnarray}
    \frac{\partial \braket{Y_i^{(o)}}}{\partial \phi} &=& \frac{\partial\braket{(C_la_i^{(i)} - C_l^*a_i^{\dagger(i)})}}{i\partial \phi} - \frac{\partial\braket{(D_l^*a_s^{(i)} - D_la_s^{\dagger(i)})}}{i\partial \phi} \nonumber \\ &=& -\sqrt{1-l_i}g_2(G_1\braket{X_s^{(i)}(\phi)} + g_1\braket{X_i^{(i)}(-\phi)}) \qquad \quad
\end{eqnarray}
The resultant equation for sensitivity is then given as
\begin{eqnarray}
    (\Delta \phi)_{Y_i} = \frac{\sqrt{|C_l|^2(\Delta Y_i^{(i)})^2 + |D_l|^2(\Delta Y_s^{(i)})^2 + 2|C_l||D_l|(\braket{Y_s^{(i)}Y_i^{(i)}} - \braket{Y_s^{(i)}}\braket{Y_i^{(i)}}) + |C_l'| + |D_l'|}}{\sqrt{1-l_i}g_2|G_1\braket{X_s^{(i)}} + g_1\braket{X_i^{(i)}}|}.
\end{eqnarray}
From the expressions above, the sensitivity for the dual port measurement, at both the bright and dark fringes, is obtained as 
\begin{eqnarray}
    \frac{\partial \braket{Y_{\pm}^{(o)}}}{\partial \phi} = (\sqrt{1-l_s}G_2 \mp \sqrt{1-l_i}g_2)(G_1\braket{X_s^{(i)}(\phi)} + g_1\braket{X_i^{(i)}(-\phi)}) \qquad \quad
\end{eqnarray}
\begin{eqnarray}
    (\Delta \phi)_{Y_{\pm}} = \frac{\sqrt{|E_l|^2(\Delta Y_s^{(i)})^2 + |F_l|^2(\Delta Y_i^{(i)})^2 + 2|E_l||F_l|(\braket{Y_s^{(i)}Y_i^{(i)}} - \braket{Y_s^{(i)}}\braket{Y_i^{(i)}}) + |E_l'| + |F_l'|}}{(\sqrt{1-l_s}G_2 \mp \sqrt{1-l_i}g_2)|G_1\braket{X_s^{(i)}} + g_1\braket{X_i^{(i)}}|}.
\end{eqnarray}

\subsubsection{Truncated interferometer}
Finally, we state the expressions for the sensitivity of the dual-port measurements for the noisy truncated interferometer. Identical to the ideal case, for the truncated interferometer, the expression for the phase change signal is given as
\begin{eqnarray}
    \frac{\partial \braket{Y_{\pm}^{(o)}}}{\partial \phi} &=& \frac{\partial \braket{Y_s^{(o)}}}{\partial \phi} = \frac{\partial\braket{(A_la_s^{(i)} - A_l^*a_s^{\dagger(i)})}}{i\partial \phi} - \frac{\partial\braket{(B_l^*a_i^{(i)} - B_la_i^{\dagger(i)})}}{i\partial \phi}  \nonumber \\ &=& \sqrt{1-l_s}(G\braket{e^{i\phi}a_s^{(i)} + e^{-i\phi}a_s^{\dagger(i)}} + g\braket{e^{-i\phi}a_i^{(i)} + e^{i\phi}a_i^{\dagger(i)}})\nonumber\\ &=& \sqrt{1-l_s}(G\braket{X_s^{(i)}(\phi)} + g\braket{X_i^{(i)}(-\phi)})
\end{eqnarray}
The resultant equation for the sensitivity at the output port for dual-port measurements at both the dark and bright fringes is given as
\begin{eqnarray}
    (\Delta \phi)_{Y_{\pm}} = \frac{\sqrt{|E_l|^2(\Delta Y_s^{(i)})^2 + |F_l|^2(\Delta Y_i^{(i)})^2 + 2|E_l||F_l|(\braket{Y_s^{(i)}Y_i^{(i)}} - \braket{Y_s^{(i)}}\braket{Y_i^{(i)}}) + |E_l'| + |F_l'|}}{\sqrt{1-l_s}|G\braket{X_s^{(i)}} + g\braket{X_i^{(i)}}|}
\end{eqnarray}

\section{Mechanisms for sensitivity enhancement in joint homodyne detection scenarios}
\label{Sensitivity enhancement}
The aim of this appendix is to shed light on the mechanisms leading to the sensitivity improvement under joint homodyne detection for both the full and the truncated interferometers. This permits us to understand whether the performance improvement is provided by an increase in the signal due to the gain and/or a reduction of the noise due to squeezing. 
\subsection{Full SU(1,1) interferometer}
In the case of joint homodyne detection in a full SU(1,1) interferometer, i. e., with two PAs, the sensitivity is given by Eqs.\,(\ref{joint dark},\ref{joint bright}). In this situation, the output noise is given by Eq. (\ref{Joint variance}). 
\subsubsection{Interferometer operation at the dark fringe}
At the dark fringe operation point $\phi_0 = (2n+1)\pi$, the variance of $Y_{-}^{(o)}$ is given by
\begin{equation}
    (\Delta Y^{(o)}_{-})^2 = \frac{(G_2 + g_2)^2}{(G_1 + g_1)^2}  \left[(\Delta Y_s^{(i)})^2 +(\Delta Y_i^{(i)})^2 + 2(\braket{Y_s^{(i)}Y_i^{(i)}} - \braket{Y_s^{(i)}}\braket{Y_i^{(i)}})\right]\ .
\end{equation}
 The term between brackets corresponds to the variance of the quadrature $Y_-^{(i)}=Y_s^{(i)}-Y_i^{(i)}$ for the input fields. Consequently,  the output noise can, at best, be equal to the input noise when the two PAs have the same gain.  Otherwise, this noise is always larger than the input noise when $G_1 \leq G_2 < \infty$.
 
Besides following Eq. (\ref{phase change signal full}), the signal response to a phase change at $\phi_0 = (2n+1)\pi$ is given by
\begin{eqnarray}
     \frac{\partial \braket{Y_{-}^{(o)}}}{\partial \phi} = (G_2 + g_2)(G_1\braket{X_s^{(i)}} + g_1\braket{X_i^{(i)}})\,.\label{Eq098}
\end{eqnarray}
Consequently, the sensitivity enhancement at this operating point is due solely to the enhancement of the signal (see Eq.\,\ref{Eq098}) because the output noise is at best the same as the total input noise. 

\subsubsection{Interferometer operation at the bright fringe}
Conversely, at the bright fringe operation point $\phi_0 = 2n\pi$, the output noise and the signal derivative with respect to the phase shift are respectively given by
\begin{eqnarray}
    (\Delta Y^{(o)}_{+})^2& = &\frac{(\Delta Y_s^{(i)})^2 +(\Delta Y_i^{(i)})^2 + 2(\braket{Y_s^{(i)}Y_i^{(i)}} - \braket{Y_s^{(i)}}\braket{Y_i^{(i)}})}{(G_1 + g_1)^2 (G_2 + g_2)^2}\ ,\label{Eq099}\\
     \frac{\partial \braket{Y_{+}^{(o)}}}{\partial \phi} &=& (G_2 - g_2)(G_1\braket{X_s^{(i)}} + g_1\braket{X_i^{(i)}}).
\end{eqnarray}
Equation (\ref{Eq099}) shows that the parametric gain reduces the output noise with respect to the total input noise. However, there is no obvious increase of the signal due to the parametric gain.

\subsection{Truncated interferometer}
We consider in this section the SU(1,1) interferometer in truncated configuration, i.e., with one PA only and a joint homodyne detection. The results are given in Eqs.\,(\ref{truc bright}) and (\ref{truc dark}). Depending on whether one operates the truncated interferometer at the dark fringe or at the bright fringe, the quadrature that provides the optimum sensitivity is either $Y_-^{(o)}$ or $Y_+^{(o)}$. Since these two situations lead to the same results for the noise variance and the derivative of the signal with respect to the phase, we give below the results only for $Y_-^{(o)}$ at the dark fringe. The expressions for $Y_+^{(o)}$ at the bright fringe are identical.

The noise variance of the joint homodyne detection is thus given by
\begin{equation}
    (\Delta Y^{(o)}_{-})^{2} = \frac{(\Delta Y_s^{(i)})^2 +(\Delta Y_i^{(i)})^2 + 2(\braket{Y_s^{(i)}Y_i^{(i)}} - \braket{Y_s^{(i)}}\braket{Y_i^{(i)}})}{(G_1 + g_1)^{2}}\ ,
\end{equation}
and the derivative with respect to the phase of the average signal is
\begin{equation}
     \frac{\partial \braket{Y_{-}^{(o)}}}{\partial \phi} = (G_1\braket{X_s^{(i)}} + g_1\braket{X_i^{(i)}})\ .
\end{equation}


Consequently, in this situation, the improvement in sensitivity arises from both a reduction of the noise variance with respect to one of the input fields and from an amplification of the average output signal.




In conclusion, although the full and the truncated SU(1,1) interferometers achieve the same sensitivity, the mechanisms through which this improvement occurs are different. 

\section{Formulary of Sensitivity Expressions}
\label{sen table}
We present here a consolidated summary of the sensitivity expressions derived in Section \ref{sen exp} for arbitrary input states. The results are organized according to interferometer configurations and measurement schemes in two tables, for the ideal and lossy interferometers, successively. The bottom parts of each table give the sensitivity expressions in the particular case of an injected coherent state. We highlight the novelty of our work giving, when relevant, previous derivations from the literature.
\subsection{Ideal Interferometer}

\begin{table}[h]
\centering
\caption{\textbf{Minimum sensitivity expressions without losses.}}
\label{table1}
\vspace{0.3cm}

\begin{tabular}{p{1.5cm} p{3cm} p{5cm} p{3.5cm} p{1.8cm}}
\hline\hline

\multicolumn{5}{c}{\textbf{Minimum sensitivity expressions without losses}} \\
\hline \\ [-8pt]

\shortstack{\textbf{Input}\\\textbf{State}} &
\shortstack{\textbf{PA}\\\textbf{Configuration}} & 
\shortstack{\textbf{Measurement}\\\textbf{Scheme}} & 
\shortstack{\textbf{Minimum}\\\textbf{Sensitivity}} & 
\textbf{Ref.} \\ [8pt]

\hline \\

$\rho_{s,i}$
& \shortstack{Balanced\\($G_1=G_2$)} 
& \shortstack{ \vspace{-0.1cm} Signal port\\ homodyne detection $Y^{(o)}_{s}$} 
& \textbf{Eq.(\ref{Eq13})}
& This work \\[8pt]

$\rho_{s,i}$
& \shortstack{Unbalanced\\($G_2\gg G_1$)} 
& \shortstack{\vspace{-0.1cm} Signal port\\ homodyne detection $Y^{(o)}_{s}$} 
& \textbf{Eq.(\ref{Eq14})}
& This work \\ [8pt]

$\rho_{s,i}$
& \shortstack{Balanced\\($G_1=G_2$)} 
& \shortstack{Joint homodyne\\ detection $Y^{(o)}_{\pm}$} 
& \textbf{Eq.(\ref{joint dark})}, \textbf{Eq.(\ref{joint bright})}
& This work \\ [8pt]

$\rho_{s,i}$
& \shortstack{Truncated\\ One PA ($G$)} 
& \shortstack{Joint homodyne\\ detection $Y^{(o)}_{\pm}$} 
& \textbf{Eq.(\ref{truc bright})}, \textbf{Eq.(\ref{truc dark})}
& This work \\[16pt]

\hline \\

$\ket{\alpha,0}_{s,i}$
& \shortstack{Balanced\\($G_1 = G_2$)} 
& \shortstack{ \vspace{-0.1cm} Signal port\\ homodyne detection $Y^{(o)}_{s}$} 
& $\frac{1}{2G^2 |\alpha|}$ 
& \cite{21_ou2020quantum} \\ [8pt]

$\ket{\alpha,0}_{s,i}$ 
& \shortstack{Unbalanced\\($G_2 \gg G_1$)} 
& \shortstack{\vspace{-0.1cm} Signal port\\ Homodyne detection $Y^{(o)}_{s}$} 
& $\frac{1}{\sqrt{2}G_1(G_1 + g_1) |\alpha|}$
& \cite{21_ou2020quantum} \\ [8pt]

$\ket{\alpha,0}_{s,i}$ 
& \shortstack{Balanced\\($G_1 = G_2$)} 
& \shortstack{Joint homodyne\\ detection $Y^{(o)}_{\pm}$} 
& $\frac{1}{\sqrt{2}G(G + g) |\alpha|}$
& \cite{21_ou2020quantum} \\ [8pt]

$\ket{\alpha,0}_{s,i}$ 
& \shortstack{Truncated\\ One PA ($G$)} 
& \shortstack{Joint homodyne\\ detection $Y^{(o)}_{\pm}$} 
& $\frac{1}{\sqrt{2}G(G + g) |\alpha|}$
& \cite{28_anderson2017phase,29_gupta2018optimized} \\ [8pt]

\hline\hline
\end{tabular}

\end{table}

\newpage
\subsection{Lossy interferometer}
\subsubsection{Internal Losses}

\hspace{-1cm}
\begin{table}[h]
\centering
\caption{\textbf{Minimum sensitivity expressions in presence of internal losses.}}
\label{table2}
\vspace{0.3cm}

\begin{tabular}{p{1.5cm} p{3cm} p{5cm} p{5cm} p{1.8cm}}
\hline\hline

\multicolumn{5}{c}{\textbf{Minimum sensitivity expressions in presence of internal losses}} \\
\hline \\ [-8pt]

\shortstack{\textbf{Input}\\\textbf{State}} &
\shortstack{\textbf{Amplifier}\\\textbf{Configuration}} & 
\shortstack{\textbf{Measurement}\\\textbf{Scheme}} & 
\shortstack{\textbf{Minimum}\\\textbf{Sensitivity}} & 
\textbf{Ref.} \\ [8pt]

\hline \\
$\rho_{s,i}$
& \shortstack{Balanced\\($G_1=G_2$)} 
& \shortstack{\vspace{-0.1cm} Signal port\\ homodyne detection $Y^{(o)}_{s}$}
&\shortstack{\textbf{Eq.(\ref{noisy_sens_1})} $(\boldsymbol{L_s \neq L_i})$\\\textbf{Eq.(\ref{noisy_1})} $(\boldsymbol{L_s = L_i})$}
& This work \\ [8pt]

$\rho_{s,i}$
& \shortstack{Unbalanced\\($G_2\gg G_1$)} 
& \shortstack{\vspace{-0.1cm} Signal port\\ homodyne detection $Y^{(o)}_{s}$} 
&\shortstack{\textbf{Eq.(\ref{noisy_sens_1})} $(\boldsymbol{L_s \neq L_i})$\\\textbf{Eq.(\ref{noisy_2})} $(\boldsymbol{L_s = L_i})$}
& This work \\ [8pt]

$\rho_{s,i}$
& \shortstack{Balanced\\($G_1=G_2$)} 
& \shortstack{Joint homodyne\\ detection $Y^{(o)}_{\pm}$} 
&\shortstack{ \textbf{Eq.(\ref{noisy_sens_j})} $(\boldsymbol{L_s \neq L_i})$\\\textbf{Eqs.(\ref{noisy_3}),(\ref{noisy_4})} $(\boldsymbol{L_s = L_i})$}
& This work \\ [8pt]

$\rho_{s,i}$
& \shortstack{Truncated\\ One PA ($G$)} 
& \shortstack{Joint homodyne\\ detection $Y^{(o)}_{\pm}$} 
&\shortstack{\textbf{Eq.(\ref{Eq29})}\\ $(\boldsymbol{L_s \neq L_i})$}
& This work \\[16pt]

\hline \\ 

$\ket{\alpha,0}_{s,i}$
& \shortstack{Balanced\\($G_1 = G_2$)} 
& \shortstack{\vspace{-0.1cm} Signal port\\ homodyne detection $Y^{(o)}_{s}$} 
&\shortstack{\textbf{Eq.(\ref{28.sph})}\\ $(\boldsymbol{L_s = L_i})$} 
& \cite{32_ou2012enhancement} \\ [8pt]

$\ket{\alpha,0}_{s,i}$ 
& \shortstack{Unbalanced\\($G_2 \gg G_1$)} 
& \shortstack{\vspace{-0.1cm} Signal port\\ Homodyne detection $Y^{(o)}_{s}$} 
&\shortstack{\textbf{Eq.(\ref{29.dhd})}\\ $(\boldsymbol{L_s = L_i})$}
& This work \\ [8pt]

$\ket{\alpha,0}_{s,i}$ 
& \shortstack{Balanced\\($G_1 = G_2$)} 
& \shortstack{Joint homodyne\\ detection $Y^{(o)}_{\pm}$} 
&\shortstack{\textbf{Eq.(\ref{29.dhd})}\\ $(\boldsymbol{L_s = L_i})$}
& This work \\ [8pt]

\hline\hline
\end{tabular}

\end{table}
\newpage

\subsubsection{External Losses}

\begin{table}[h]
\centering
\caption{\textbf{Minimum sensitivity expressions in presence of external losses.}}

\label{table3}
\vspace{0.3cm}

\begin{tabular}{p{1.5cm} p{3cm} p{5cm} p{5cm} p{1.8cm}}
\hline\hline

\multicolumn{5}{c}{\textbf{Minimum sensitivity expressions in presence of external losses}} \\
\hline \\ [-8pt]

\shortstack{\textbf{Input}\\\textbf{State}} &
\shortstack{\textbf{Amplifier}\\\textbf{Configuration}} & 
\shortstack{\textbf{Measurement}\\\textbf{Scheme}} & 
\shortstack{\textbf{Minimum}\\\textbf{Sensitivity}} & 
\textbf{Ref.} \\ [8pt]

\hline \\

$\rho_{s,i}$
& \shortstack{Balanced\\($G_1=G_2$)} 
& \shortstack{\vspace{-0.1cm} Signal port\\ homodyne detection $Y^{(o)}_{s}$}
&\shortstack{\textbf{Eq.(\ref{external1})} \\\textbf{Eq.(\ref{corr1})} }
& This work \\ [8pt]

$\rho_{s,i}$
& \shortstack{Unbalanced\\($G_2\gg G_1$)} 
& \shortstack{\vspace{-0.1cm} Signal port\\ homodyne detection $Y^{(o)}_{s}$} 
&\shortstack{\textbf{Eq.(\ref{external1})} \\\textbf{Eq.(\ref{corr2})} }
& This work \\ [8pt]

$\rho_{s,i}$
& \shortstack{Balanced\\($G_1=G_2$)} 
& \shortstack{Joint homodyne\\ detection $Y^{(o)}_{\pm}$} 
&\shortstack{ \textbf{Eq.(\ref{external2})} $(\boldsymbol{l_s \neq l_i})$\\\textbf{Eqs.(\ref{corr3}),(\ref{corr4})} $(\boldsymbol{l_s = l_i})$}
& This work \\ [8pt]

\hline\\

$\ket{\alpha,0}_{s,i}$
& \shortstack{Balanced\\($G_1 = G_2$)} 
& \shortstack{\vspace{-0.1cm} Signal port\\ homodyne detection $Y^{(o)}_{s}$} 
& \shortstack{$\frac{\sqrt{1 + \frac{l}{1-l}}}{2G^2|\alpha|}$\\(independent of $l_i$)} 
& \cite{32_ou2012enhancement} \\ [8pt]

$\ket{\alpha,0}_{s,i}$ 
& \shortstack{Unbalanced\\($G_2 \gg G_1$)} 
& \shortstack{\vspace{-0.1cm} Signal port\\ homodyne detection $Y^{(o)}_{s}$} 
& \shortstack{$\frac{\sqrt{1+\frac{l}{2(1-l)}\left(\frac{G_1+g_1}{G_2}\right)^2\ }
}{\sqrt{2}G_1(G_1+g_1)|\alpha|}\,$ \\(independent of $l_i$)}
& This work \\ [8pt]

$\ket{\alpha,0}_{s,i}$ 
& \shortstack{Balanced\\($G_1 = G_2$)} 
& \shortstack{Joint homodyne\\ detection $Y^{(o)}_{-}$} 
&\shortstack{$\frac{\sqrt{1+\frac{l}{(1-l)}}}{\sqrt{2}G_1(G_1+g_1)|\alpha|}$ $(\boldsymbol{l_s = l_i})$}
& This work \\ [8pt]

$\ket{\alpha,0}_{s,i}$ 
& \shortstack{Balanced\\($G_1 = G_2$)} 
& \shortstack{Joint homodyne\\ detection $Y^{(o)}_{+}$} 
&\shortstack{$\frac{\sqrt{1+\frac{l}{(1-l)}{(G_1 + g_1)^4}}}{\sqrt{2}G_1(G_1+g_1)|\alpha|}$ $(\boldsymbol{l_s = l_i})$}
& This work \\ [8pt]

\hline\hline
\end{tabular}

\end{table}

\begin{figure}[htbp]
\begin{subfigure}{0.50\columnwidth}
\centering
\includegraphics[width=\linewidth]{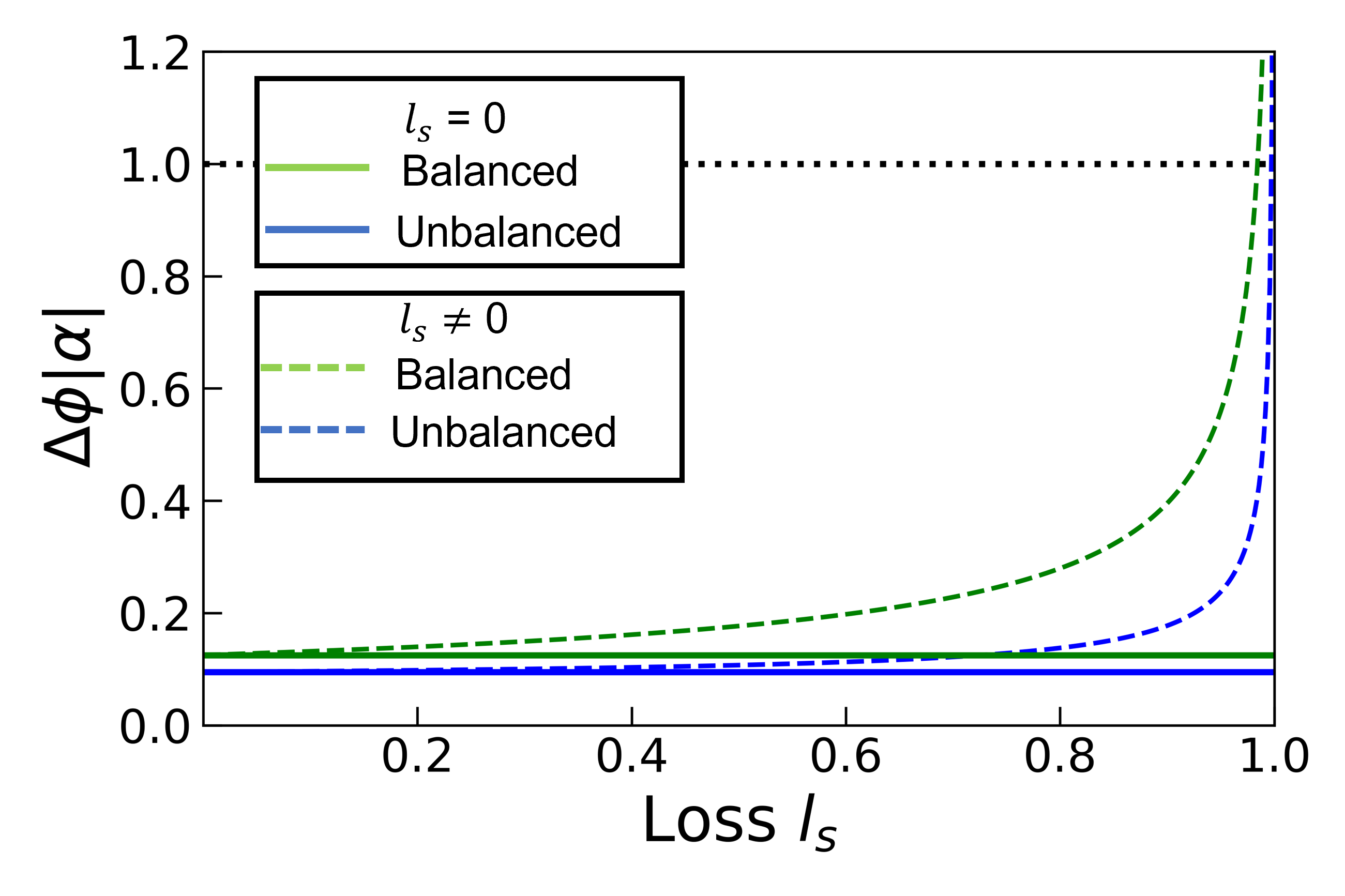}
\caption{}
\end{subfigure}
\hfill
\begin{subfigure}{0.50\columnwidth}
\centering
\includegraphics[width=\linewidth]{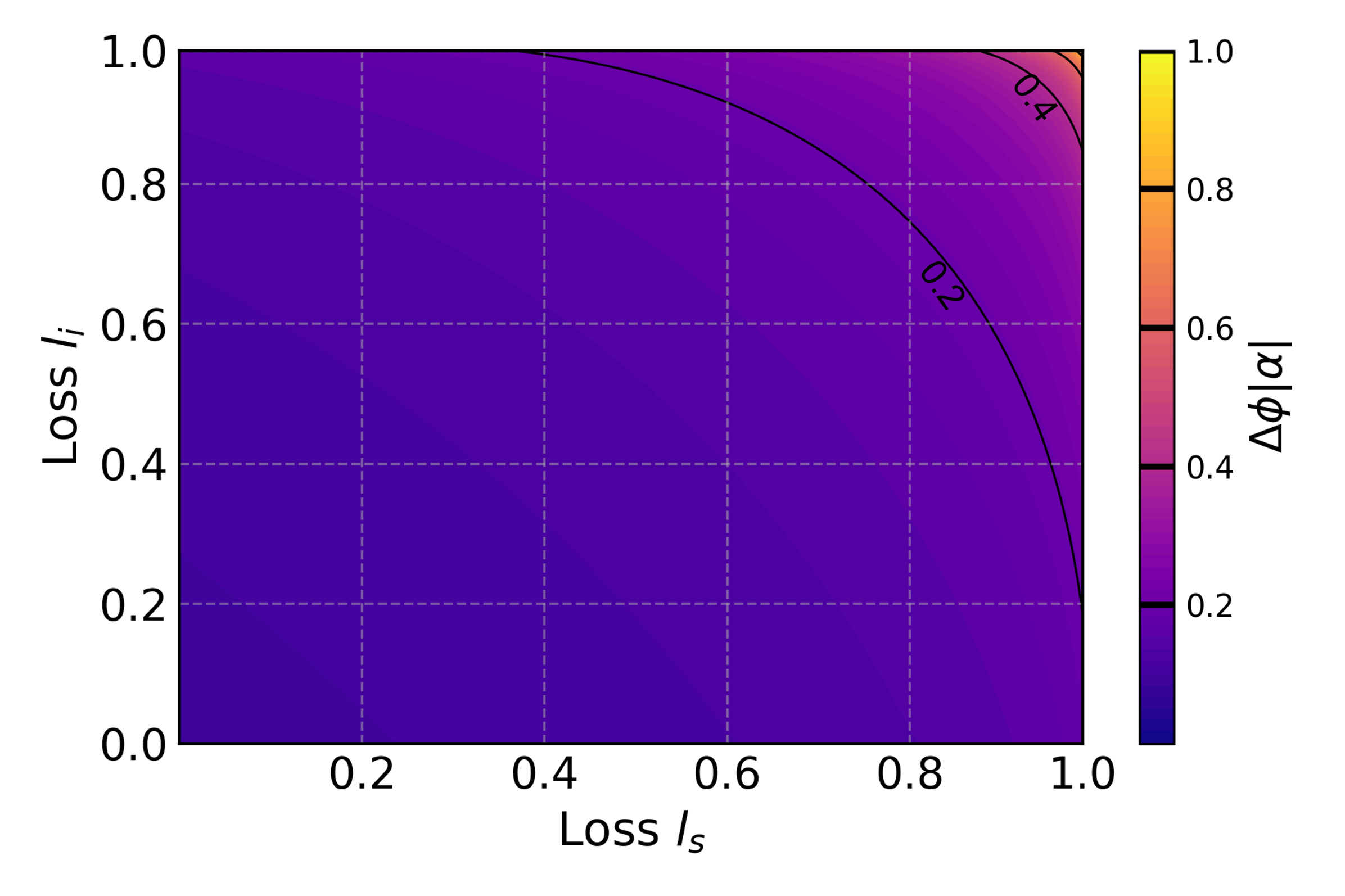}
\caption{}
\end{subfigure}
\vspace{0.5cm}
\begin{subfigure}{0.55\columnwidth}
\centering
\includegraphics[width=\linewidth]{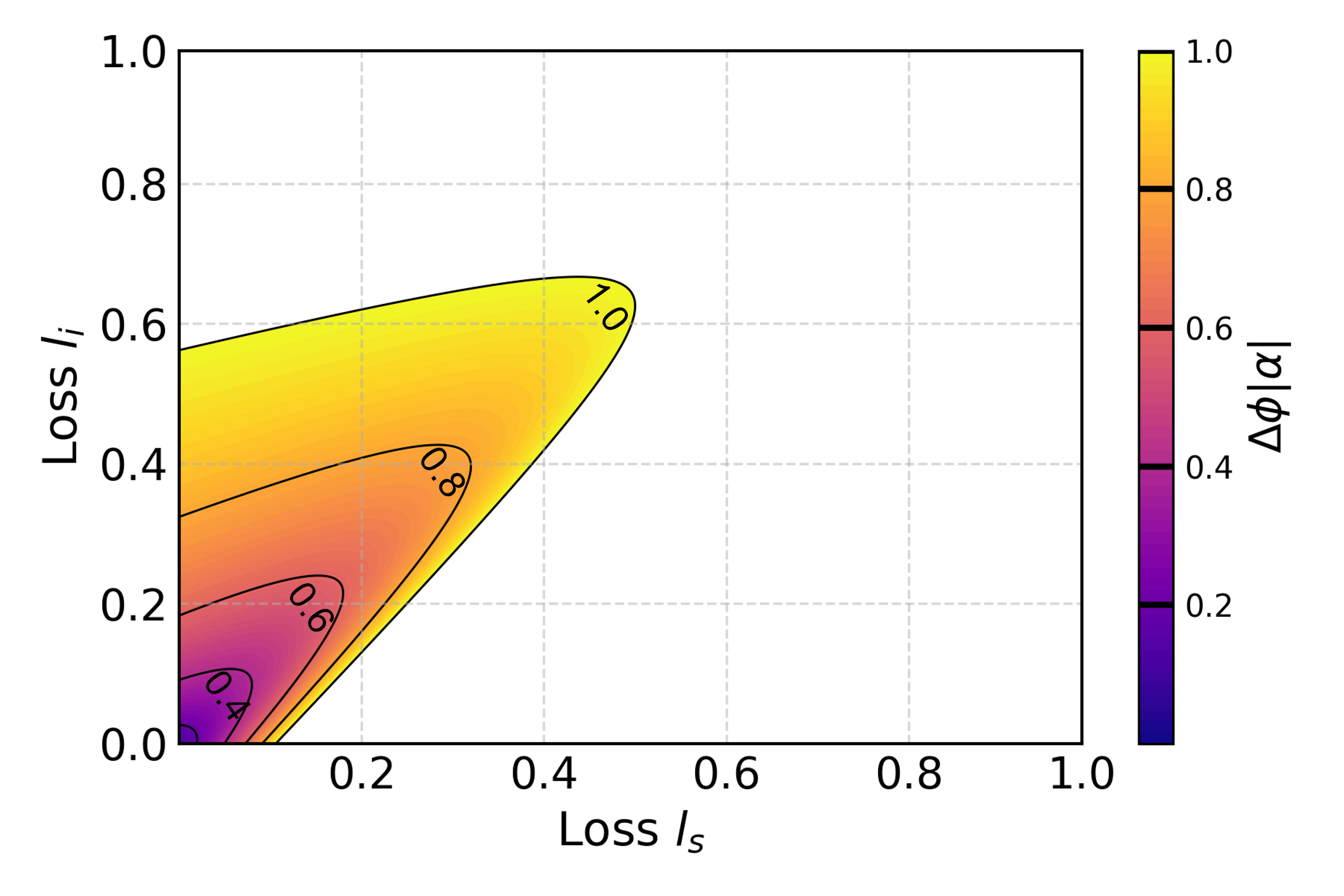}
\caption{}
\end{subfigure}

\caption{Variation of phase sensitivity as a function of external losses in the two output ports of the SU(1,1) interferometer. (a) Single-port detection with balanced ($G_1=G_2=2$) and unbalanced gains ($G_1=2$, $G_2=5$) of the two PAs. Notably, the external loss in the other output has no effect. (b) Joint homodyne detection at dark fringe point, corresponding to $Y_{-}$ quadrature, with balanced gain ($G_1=G_2=2$). (c) Joint homodyne detection at bright fringe point, corresponding to $Y_{+}$ quadrature, with balanced gain ($G_1=G_2=2$).}
\label{fig: external_loss}
\end{figure}

\end{document}